\documentclass[prb,aps,twocolumn,superscriptaddress]{revtex4-2}
\usepackage{graphicx}
\usepackage{dcolumn}
\usepackage{bm}
\usepackage{lipsum}
\usepackage{ragged2e}
\usepackage{soul}
\usepackage{braket}
\usepackage{bbm}
\usepackage{lipsum}
\usepackage{color}
\usepackage{xcolor}
\usepackage{amsfonts}
\usepackage{amsmath}
\usepackage{amsmath,amssymb,latexsym}
\usepackage{wasysym,stmaryrd,marvosym}
\usepackage{mathrsfs}
\usepackage{dsfont}
\usepackage{float}
\providecommand{\keywords}[1]{\textbf{\textit{Index terms---}} #1}
\usepackage[mathlines]{lineno}
\usepackage[colorinlistoftodos]{todonotes}
\usepackage[colorlinks=true, allcolors=blue]{hyperref}
\usepackage{array}

\begin{document}
\newcommand{\cnyn}{Centro de Nanociencias y Nanotecnolog\'ia,
Universidad Nacional Aut\'onoma de M\'exico, Apartado Postal 2681, 22800, Ensenada, Baja California, M\'exico.}
\newcommand{\uabc}{Facultad de Ciencias, Universidad Aut\'onoma de Baja California, Apartado Postal 1880, 22800 Ensenada, Baja California, M\'exico}
\newcommand{\ou}{Department of Physics and Astronomy and Nanoscale and Quantum Phenomena Institute, Ohio University, Athens, Ohio 45701}

\title{Thermal difference reflectivity of tilted 2D Dirac materials}

\author{M. A. Mojarro}
\affiliation{\ou}
\author{R. Carrillo-Bastos}
\affiliation{\uabc}
\author{Jes\'us A. Maytorena}
\email{jesusm@ens.cnyn.unam.mx}
\affiliation{\cnyn}

\date{\today}

\begin{abstract}
Deviation from perfect conical dispersion in Dirac materials, such as the presence of mass or tilting, enhances control and directionality of electronic transport. To identify these signatures, we analyze the thermal derivative spectra of optical reflectivity in doped massive tilted Dirac systems. The density of states and chemical potential are determined as preliminary steps to calculate the optical conductivity tensor at finite temperature using thermal convolution. Changes in reflection caused by temperature variations enable clear identification of critical frequencies in the optical response. By measuring these spectral features in the thermoderivative spectrum, energy gaps and band structure tilting can be determined. A comparison is presented between the spectra of various low-energy Dirac Hamiltonians. Our findings suggest that thermal difference spectroscopy holds promise as a valuable technique for probing interband transitions of 2D Dirac fermions

\end{abstract}


\maketitle

Optical spectroscopies allow investigating surfaces and interfaces, offering non-destructive, in-situ, and real-time probing capabilities, with various techniques available such as spectroscopic ellipsometry, differential reflectance, electroreflectance, reflection anisotropy spectroscopy,
magneto-optics, and nonlinear spectroscopies \cite{PhotoProbe,ProbSurf}. Recent methods resolve single nano-objects and subwavelength structures \cite{novotny2012principles},
 plasmon dynamics \cite{Downer2016}, 
 organic and biological interfaces \cite{Zhang2013},
 graphene-like materials \cite{Geoffroy2016},
 and metasurfaces \cite{Sarychev2022,Chang2018}. 
 
 In the field of 2D systems, the in-plane optical anisotropy of low-symmetry 2D materials is emerging as a unique characteristic with potential applications in optics, optoelectronics, and photonics \cite{HuangOpt2D,MaTunable,Gan2Dmat}. 
Orthorhombic systems like black phosphorus (BP) \cite{XinBP,WangBP} 
and group IV monochalcogenides, monoclinic systems like 1T$'$-WTe$_2$ \cite{ZhangWTe2}
and triclinic materials like ReS$_2$ \cite{ShenReS2}
present an optical anisotropy due to their anisotropic band structure, which can be modified by band engineering methods. This is an essential difference with 
anisotropic nanostructures made of symmetric 2D materials, where the anisotropy of the dielectric constant is designed by nanofabrication \cite{LiEmerg}.
  
The in-plane anisotropy of these materials, along with critical frequencies and optical transitions involving the Fermi level, have been probed through a number of differential and modulation spectroscopy techniques, like differential reflectance and transmittance spectroscopy \cite{FrisendaMicro},
anisotropic optical absorption and photoluminescence \cite{LiEmerg}, 
Raman scattering \cite{LiEmerg,ZhouOpt},
time-domain thermoreflectance \cite{SunTemp},
azimuthal-dependent reflectance difference spectroscopy \cite{ShenResolv},
or reflection difference microscopy combined with electron transport measurements and atomic force microscopy \cite{TaoMech}.

2D materials with tilted Dirac cones in their energy spectrum are an interesting and currently attractive variant of
anisotropic systems. The tilting of the bands introduces an additional source
of anisotropy which can modify significantly the optical response 
\cite{verma,seminal,plasmonsletter,TanLifs,WildOptical,TanAnisot,AntoniosPolariz,Fu2023,Park2022}.
The quasi-2D organic conductor $\alpha$-(BEDT-TTF)$_2$I$_3$ \cite{Uykur2019,Hirata2016,HirataAnomalous,OhkiChiral,OhkiGap} is a well known example 
of an anisotropic material which presents a pair of tilted Dirac cones when external pressure is applied \cite{MesslessFermions,OsadaCurrentInduced}, and even massive Dirac fermions below a critical pressure \cite{osadaExp}. Another well-studied
tilted system is the 8-$Pmmn$ borophene, for which band gap opening  \cite{Wang2019}
and tilt-tuning \cite{Yekta2023}
have been predicted.

Herein, we explore theoretically the thermal difference spectroscopy \cite{Holcomb,Holcomb1994,Holcomb1996}
to identify relevant
features in the optical response of massive tilted 2D Dirac systems at finite 
temperature. To this end, we first calculate the temperature dependence of the chemical potential and
then the optical conductivity through a thermal convolution. 
This extends our previously reported calculations at zero temperature \cite{seminal,plasmonsletter}.
Then we evaluate the optical reflectivity at two close temperatures to obtain
the derivative of its spectrum, which provides a way to compensate for the reduced structure
produced by thermal broadening.
The change in reflectivity caused by temperature 
variation probes the anisotropy of the system and highlights critical points 
in the optical absorption or reflection spectrum. Thermal difference spectroscopy
measures the derivative of the optical spectrum, as in temperature modulation spectroscopy,
but without involving modulation of the sample's temperature at a given frequency.
In particular, we discuss the possibility to 
estimate parameters like tilting or gaps through this thermal derivative approach.

From the electromagnetic scattering problem of optical reflection and refraction at a flat interface 
made of a 2D system, with conductivities $\sigma_{xx}(\omega,T)$ and $\sigma_{yy}(\omega,T)$,
separating two homogeneous media with dielectric constants $\epsilon_1$ and $\epsilon_2$, 
the optical reflectivity is obtained as $R(\omega,T)=|r_p|^2\cos^2\phi+|r_s|^2\sin^2\phi$
 \cite{seminal}, where $\phi$ is the angle of polarization of the incident field, $\omega$
 the frequency, and $T$ the temperature. The Fresnel
 amplitudes are
 \begin{eqnarray}
\nonumber r_p(\omega,T) &=& \frac{\epsilon_2k_z^i-\epsilon_1k_z^t+4\pi (k_z^ik_z^t/k_0)(\sigma_{xx}(\omega,T)/c)}
{\epsilon_2k_z^i+\epsilon_1k_z^t+4\pi (k_z^ik_z^t/k_0)(\sigma_{xx}(\omega,T)/c)}\ ,  \\ \\
r_s(\omega,T) &=& \frac{k_z^i-k_z^t-4\pi k_0(\sigma_{yy}(\omega,T)/c)} 
{k_z^i+k_z^t+4\pi k_0(\sigma_{yy}(\omega,T)/c)} \ ,
\end{eqnarray}
for $p$ ($\phi=0$) and $s$ ($\phi=\pi/2$) polarizations; $k_0=\omega/c$, and $k_z^i=k_0\sqrt{\epsilon_1}\cos\theta_i$,  $k_z^t=k_0\sqrt{\epsilon_2-\epsilon_1\sin^2\theta_i}$, are the normal to the surface components of the incident
and refracted wave vectors, respectively, where $\theta_i$ is the incidence angle.
The frequency and temperature dependence of $R_p$ ($R_s$) is determined solely
by $\sigma_{xx}$ ($\sigma_{yy}$).


In the following we take the thermal derivative of the reflectivity spectra. To this end we
calculate the normalized difference between $R(T)$ taken at two temperatures 
\cite{Holcomb,Holcomb1996} 
\begin{displaymath}
\frac{R_{\nu}(\omega,T+\Delta T)-R_{\nu}(\omega,T-\Delta T)}
{[R_{\nu}(\omega,T+\Delta T)+R_{\nu}(\omega,T-\Delta T)]/2} \,.
\end{displaymath}
Assuming a small enough $\Delta T$, the quantity
$\Delta R_{\nu}/R_{\nu}=(1/R_{\nu})(\partial R_{\nu}/\partial T)2\Delta T$ measures the change in reflectivity
caused by temperature variation. The thermoderivative is obtained from
\begin{equation}
 \frac{\partial R_{\nu}}{\partial T}=\frac{\partial R_{\nu}}{\partial \sigma'_{ii}}
 \frac{\partial \sigma'_{ii}}{\partial T}+\frac{\partial R_{\nu}}{\partial \sigma''_{ii}}
 \frac{\partial \sigma''_{ii}}{\partial T} \,,
\end{equation}
where $i=x\,(y)$ if $\nu=p\,(s)$, $\sigma'_{ii}\equiv\text{Re}(\sigma_{ii})$,
$\sigma''_{ii}\equiv\text{Im}(\sigma_{ii})$. We will show results for $p$ polarization
only ($R=R_p$), the corresponding spectra for $\nu=s$ are qualitatively similar.

The conductivity tensor at finite temperature can be calculated from a convolution
 integral between the zero temperature counterpart, $\sigma_{ij}(\omega;0,\mu')$,  
 and the peaked function $\partial f(\mu,\mu')/\partial\mu'$,
 with $f(\mu(T),\mu')=\{\exp[\beta(\mu(T)-\mu')]+1\}^{-1}$,
 which takes the form \cite{maldague}
\begin{equation} \label{sigmaT}
\sigma_{ij}(\omega;T,\mu(T))=\frac{\beta}{4}\int_{-\infty}^{\infty}\,
\frac{\sigma_{ij}(\omega;0,\mu')\,d\mu'}{\cosh^2[\beta(\mu(T)-\mu')/2]}\,,
\end{equation}
where $\mu'$ is the Fermi energy $\varepsilon_F=\mu(T=0)$, and $\beta=1/k_BT$. 
%
In this work, the zero temperature response $\sigma_{ij}(\omega;0,\mu')$ is evaluated within the Kubo formalism for a massive tilted Dirac system modeled
by the time-reversal symmetric Hamiltonian \cite{seminal}
\begin{equation} \label{H1}
H_{\xi}({\bf k})=\xi(\hbar v_tk_y\hat{\sigma}_0+\hbar v_xk_x\hat{\sigma}_x+\xi\hbar v_yk_y\hat{\sigma}_y)
+\Delta\hat{\sigma}_z \ ,
\end{equation}
with energy spectrum $\varepsilon^{\xi}_{\lambda}(k_x,k_y)=\xi\hbar v_tk_y+
\lambda\sqrt{(\hbar v_x)^2k_x^2+(\hbar v_y)^2k_y^2+\Delta^2}$, where
the Pauli matrices $\hat{\sigma}_i$ act on a pseudospin space, 
${\bf k}=(k_x,k_y)$ is the electron wave vector in the vicinity of the $K(K')$ point in the valley
$\xi=+\,(-)$, while $\lambda=\pm$ specifies the helicity of states in the conduction ($+$)
and valence ($-$) bands; for the mass term we take $\Delta>0$. A thorough study of the optical properties of 
this model (\ref{H1}) was presented in Refs.\,\onlinecite{seminal,plasmonsletter}.
This Dirac model describes several anisotropic 2D Dirac fermions in systems like 
the organic conductor $\alpha$-(BEDT-TTF)$_2$I$_3$ \cite{MesslessFermions,OsadaCurrentInduced}, 
the 8-$Pmmn$ borophene 
(with $v_x=0.86\times 10^6\,$m/s, $v_y=0.69\times 10^6\,$m/s, $v_t=0.32\times 10^6\,$m/s)
\cite{verma}, monolayer WTe$_2$ with $(v_x=0.644\times 10^6\,$m/s, $v_y=0.365\times 10^6\,$m/s, $v_t=0.464\times 10^6\,$m/s) \cite{sciPostWTe2}, or 2D ladder polyborane
($v_x=0.735\times 10^6\,$m/s, $v_y=0.397\times 10^6\,$m/s, $v_t=0.191\times 10^6\,$m/s) \cite{Fu2023}.
It has also been used in studies of nonlinear optical 
response like the nonlinear Hall effect \cite{juriNLHE,disorderNLHE,Du2018}, the second-order conductivity induced by 
the quantum metric dipole \cite{QMDCulcer1, QMDCulcer2}, and nonlinear thermal Hall effects \cite{NLThermal}.

An interesting feature of the model is that
the band gap in each valley is indirect, with a minimum (maximum) of the conduction (valence) band 
$\varepsilon^{\xi}_+$ ($\varepsilon^{\xi}_-$)
at ${\bf k}_{\xi}=-\xi Q{\bf \hat{y}}$ ($+\xi Q{\bf \hat{y}}$), where $\hbar v_yQ=\gamma\Delta/\sqrt{1-\gamma^2}$,
with $\gamma=v_t/v_y\,(0\leqslant\gamma<1)$ being the tilting parameter. As a consequence, a new region appears for the Fermi level,
$\tilde{\Delta}\leqslant\varepsilon_F\leqslant\Delta$ (the ``indirect zone''), with $\tilde{\Delta}=
\Delta\sqrt{1-\gamma^2}$, which has a striking effect on the spectrum of interband transitions
\cite{seminal}. We note that the optical response to a long wavelength external field excludes interband transitions
with finite momentum, in particular those with wave vector close to $\pm\xi 2Q{\bf\hat{y}}$, in the vicinity of the
gap $2\tilde{\Delta}$. However, the indirect gap manifests itself through the appearance of the indirect zone,
which is absent for the untilted or ungapped system ($\gamma\Delta=0$).

The evaluation of integral (\ref{sigmaT}) requires prior knowledge of the chemical potential $\mu$ as a function of temperature. Here, we obtain $\mu(T)$ by solving the transcendental equation that arises from expressing the doping electron density $n$ 
as a proper integral of the density of states (DOS), and equating $n$ to that at zero
temperature for a given Fermi energy \cite{gorbarLi2}. This approach was used in Ref.\,\onlinecite{gumbs2017}, where the function $\mu(T)$ for several doped and gapped Dirac materials (graphene, silicene, germanene, and  MoS$_2$) with a linear density of states was derived. In our work, tilting of the bands in Eq.~(\ref{H1}) is an additional component in the calculations.

Thus, to proceed, we first calculate the DOS of our system
\begin{equation}\label{DOS_3}
D(\varepsilon)=g_s\sum_{\xi,\lambda=\pm}\int\!\frac{d^2k}{(2\pi)^2}\,
\delta(\varepsilon-\varepsilon^{\xi}_{\lambda}({\bf k}))\,,
\end{equation}
with $g_s=2$ being the spin degeneracy and $\delta(\varepsilon)$ the Dirac delta function.
{
\extrarowheight = -0.4ex
\renewcommand{\arraystretch}{2.25}
\begin{table}[]
    \centering
    \begin{tabular}{||c|c|c||}
    \hline
      system & $D(\varepsilon)$ & $n$ \\ \hline\hline
        $\Delta=0$, $\gamma=0$ & $\frac{2|\varepsilon|}{\pi(\hbar v_F)^2}$ & $\frac{\varepsilon_F^2}{\pi(\hbar v_F)^2}$ \\ \hline
        $\Delta\neq0$, $\gamma=0$ & $\frac{2|\varepsilon|\Theta(|\varepsilon|-\Delta)}{\pi(\hbar v_F)^2}$ & $\frac{\varepsilon_F^2-\Delta^2}{\pi(\hbar v_F)^2}$ \\\hline
        $\Delta=0$, $\gamma\neq0$ & $\frac{2|\varepsilon|}{\pi(\hbar v_x)(\hbar v_y)(1-\gamma^2)^{3/2}}$ & $\frac{\varepsilon_F^2}{\pi(\hbar v_x)(\hbar v_y)(1-\gamma^2)^{3/2}}$\\\hline
        $\Delta\neq0$, $\gamma\neq0$ & $\frac{2|\varepsilon|\Theta(|\varepsilon|-\tilde{\Delta})}{\pi(\hbar v_x)(\hbar v_y)(1-\gamma^2)^{3/2}}$ & $\frac{\varepsilon_F^2-\Tilde{\Delta}^2}{\pi(\hbar v_x)(\hbar v_y)(1-\gamma^2)^{3/2}}$\\ \hline
    \end{tabular}
    \caption{Density of states $D(\varepsilon)$ and doping electron density 
    $n$ in terms of the Fermi energy
    $\varepsilon_F=\mu(0)$, for gapped and/or tilted Dirac systems.}
    \label{table1}
\end{table}
}
\begin{figure}
    \centering
    \includegraphics[scale=0.39]{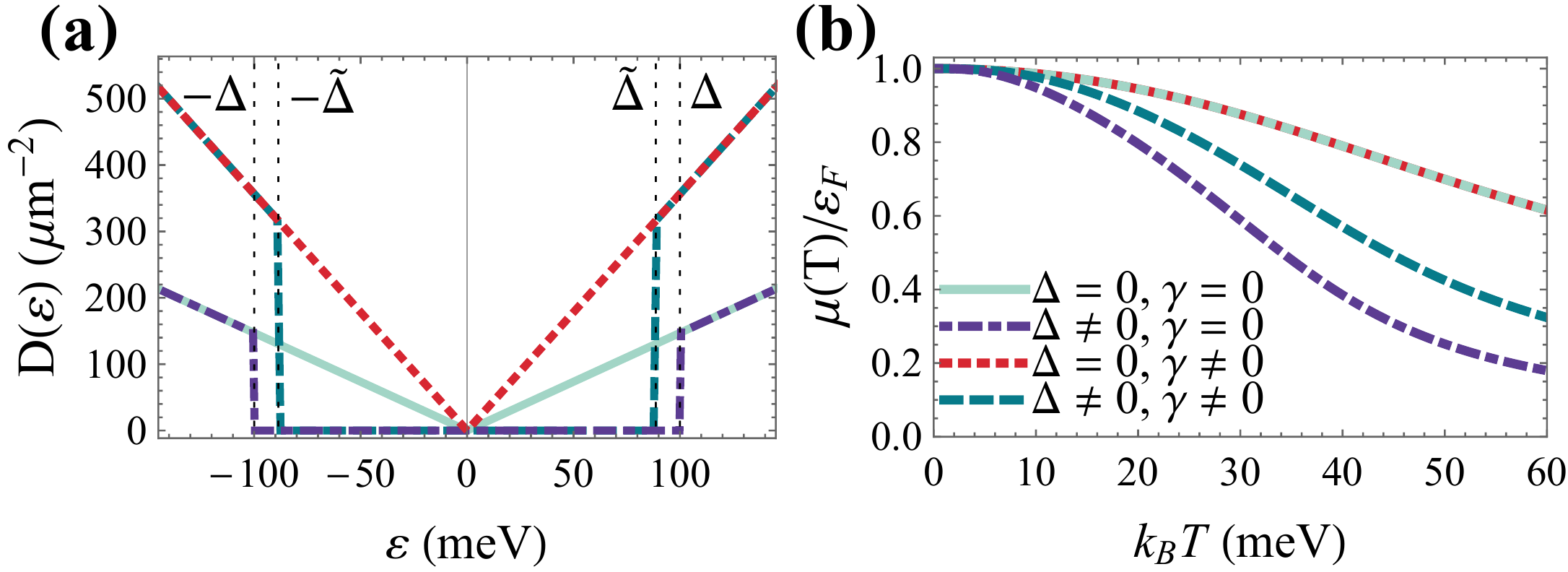}
    \caption{(a) Density of states and (b) chemical potential $\mu(T)$ of several Dirac materials: graphene ($v_x=v_y, \gamma=\Delta=0$), 
    gapped graphene ($v_x=v_y,\gamma=0, \Delta\neq 0$), tilted (8-$Pmmn$ borophene, $v_x\neq v_y, \gamma\neq 0, \Delta=0$), 
    and massive tilted ($v_x\neq v_y, \gamma\neq 0, \Delta\neq 0$) with $\varepsilon_F=110$ meV. For the gapped cases we take $\Delta=100$ meV.
    }\label{fig1}
\end{figure}
We find that the corresponding DOS of system (\ref{H1}) is electron-hole symmetric and therefore has the following structure
\begin{equation} \label{DOS_4}
D(\varepsilon)=F(\varepsilon)\Theta(|\varepsilon|-\Delta)+G(\varepsilon)\Theta(1-|\eta(\varepsilon)|)\,,
\end{equation}
where $\eta(\varepsilon)=[|\varepsilon|-(\Delta+\tilde{\Delta})/2]/[(\Delta-\tilde{\Delta})/2]$.
The function $G$ contributes when the energy is within the indirect zone $|\eta(\varepsilon)|<1$, while the function $F$ when $|\varepsilon|>\Delta$.

For $|\varepsilon|>\Delta$, the contribution $F(\varepsilon)$ reduces to the expression
\begin{equation}
F(\varepsilon)=\frac{g_sg_v}{2\pi}\frac{|\varepsilon|}{(\hbar v_F)^2}
\frac{d(\gamma I_F)}{d\gamma}\,,
\end{equation}
where $g_v=2$ is the valley degeneracy and $I_F$ is the dimensionless integral
\begin{equation}
I_F=\frac{1}{2\pi}\int_0^{2\pi}\!\frac{d\theta}{g^2(\theta)-h^2(\theta)}. \nonumber  
\end{equation}
We define the function $h(\theta)=(v_t/v_F)\sin\theta$ to describe the anisotropy resulting from
the tilt of the bands, while
$g(\theta)=[(v_x/v_F)^2\cos^2\theta+(v_y/v_F)^2\sin^2\theta]^{1/2}$ accounts for the anisotropy of the velocity, where $v_x\neq v_y$. Complex integration gives
$I_F=(v_F^2/v_xv_y)(1-\gamma^2)^{-1/2}$, which implies
\begin{equation} \label{DOS_Verma}
 F(\varepsilon)=\frac{g_sg_v}{2\pi}\frac{|\varepsilon|}{(\hbar v_x)(\hbar v_y)}
 \frac{1}{(1-\gamma^2)^{3/2}} \,.
\end{equation}
When $\tilde{\Delta}<|\varepsilon|<\Delta$ ($|\eta(\varepsilon)|<1$),
$G(\varepsilon)$ can be writen as
\begin{equation} \label{G_2}
G(\varepsilon)=\frac{g_sg_v}{2\pi}\frac{1}{(\hbar v_F)^2}\frac{\partial I_G(\varepsilon)}{\partial\varepsilon} \ ,
\end{equation}
where
\begin{equation} \label{tI}
I_G(\varepsilon) = \frac{\varepsilon}{\pi}\int_{\frac{\pi}{2}-\theta^*}^{\frac{\pi}{2}+\theta^*}\!\!d\theta\,
\frac{h(\theta)\sqrt{\varepsilon^2g^2(\theta)-\Delta^2[g^2(\theta)-h^2(\theta)]}}{[g^2(\theta)-h^2(\theta)]^2}\ ,  \nonumber
\end{equation}
with $\tan\theta^*(\varepsilon)=(v_y/v_x)\sqrt{[\varepsilon^2-\tilde{\Delta}^2]/[\Delta^2-\varepsilon^2]}$.
The restricted sector of integration reflects the drastic reduction
of the momentum space available for vertical transitions when the energy lies within the narrow stripe
between $\tilde{\Delta}$ and $\Delta$, arising from the indirect nature of the gap \cite{seminal} ($\gamma\Delta\neq 0$). 
We obtain
$I_G(\varepsilon)=\text{sign}(\varepsilon)(\varepsilon^2-\tilde{\Delta}^2)/[2(v_x/v_F)(v_y/v_F)(1-\gamma^2)^{3/2}]$,
which leads to the result $G(\varepsilon)=F(\varepsilon)$. Therefore, from (\ref{DOS_4})
we find for the DOS of a tilted and gapped Dirac system
\begin{equation} \label{DOSfinal}
D(\varepsilon)=\frac{g_sg_v}{2\pi}\frac{|\varepsilon|}{(\hbar v_x)(\hbar v_y)}
 \frac{\Theta(|\varepsilon|-\tilde{\Delta})}{(1-\gamma^2)^{3/2}} \,.
\end{equation}

\begin{figure}
    \centering
    \includegraphics[scale=0.63]{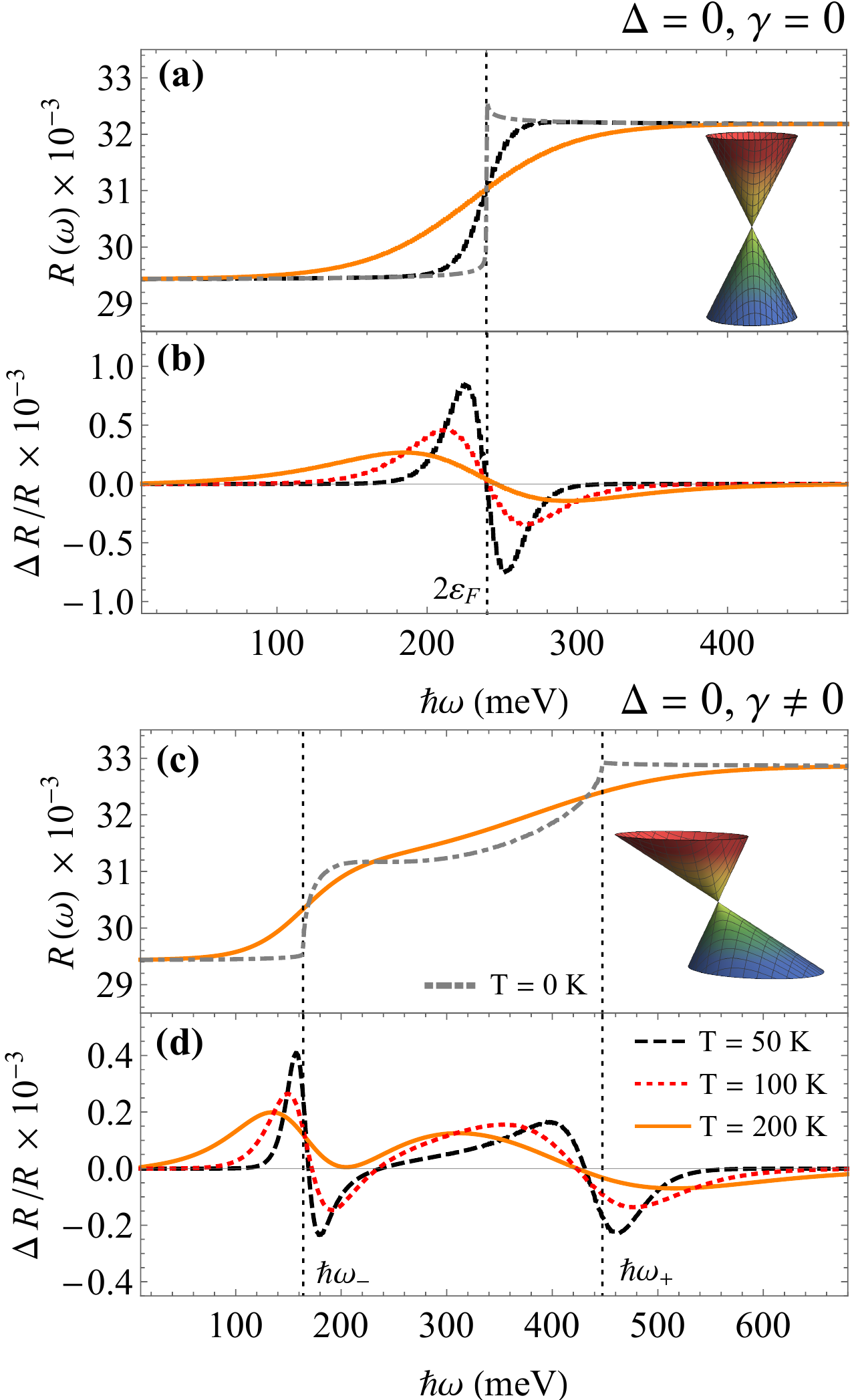}
    \caption{Optical reflectivity $R(\omega)$ at $T=200\,$K and its thermal derivative $\Delta R/R$ for several temperatures of the sample at normal incidence, corresponding to (a)-(b) graphene and (c)-(d) a gappless tilted anisotropic system (borophene 8-$Pmmn$). The insets illustrate the energy bands in each case.
    We use $\epsilon_1=1, \epsilon_2=2$, $\varepsilon_F=120\,$meV and $\Delta T=1\,$K.}
    \label{Fig2}
\end{figure}

Given that the DOS is an even function of energy, the doping electron density, obtained from the difference 
between the densities of electrons and holes \cite{gumbs2017,spikes}, can be calculated from
\begin{equation}
n(T,\mu)=\sinh(\beta\mu)\int_0^{\infty}d\varepsilon\,\frac{D(\varepsilon)}{\cosh(\beta\varepsilon)+\cosh(\beta\mu)}\,.    
\end{equation}
As a result, we find
\begin{equation} \label{nF}
n=\frac{2}{\pi (\hbar v_x)(\hbar v_y)}\frac{1}{(1-\gamma^2)^{3/2}}\frac{1}{\beta^2}\Lambda(T,\mu(T),\tilde{\Delta}) \,,
\end{equation}
where
\begin{eqnarray} 
\Lambda(T,\mu(T),\tilde{\Delta})&=&\beta\tilde{\Delta}\ln\!\left(\frac{1+e^{\beta(\mu-\tilde{\Delta})}}{1+e^{-\beta(\mu+\tilde{\Delta})}}\right) 
\nonumber \label{HT} \\
&& \hspace*{0.4cm}+\,\text{Li}_2(-e^{-\beta(\mu+\tilde{\Delta})})-\text{Li}_2(-e^{\beta(\mu-\tilde{\Delta})})\,, \nonumber
\end{eqnarray}
$\text{Li}_2(x)$ being the dilogarithm  function \cite{gorbarLi2,gradshteynLi2}. 

\begin{figure}
    \centering
    \includegraphics[scale=0.43]{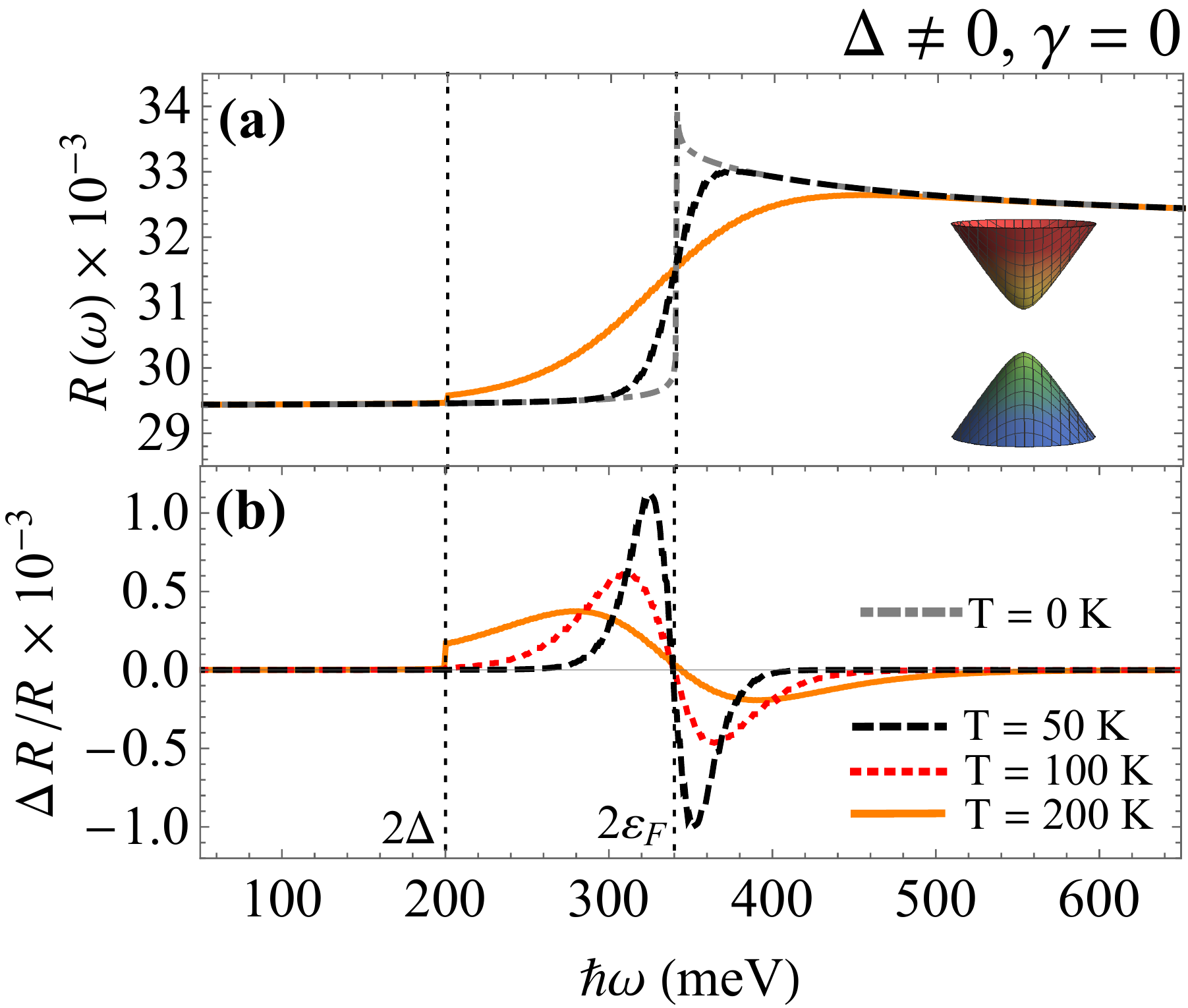}
    \caption{(a) Optical reflectivity $R(\omega)$ at $T=200\,$K and (b) its thermal derivative $\Delta R/R$ for several temperatures of the sample, corresponding to gapped graphene, with $\varepsilon_F>\Delta$. We take
    $\epsilon_2=2\epsilon_1=2$,
    $\varepsilon_F=170\,$meV, $\Delta=100\,$meV, and $\Delta T=1\,$K.}
    \label{Fig3}
\end{figure}

For fixed $n$ and $T$, the function $\mu(T)$ can be obtained from this expression or, in terms
of the Fermi energy $\varepsilon_F=\mu(0)$,
\begin{equation} \label{trascendental}
\varepsilon_F^2-\tilde{\Delta}^2=2(k_BT)^2\Lambda(T,\mu(T),\tilde{\Delta})\,. 
\end{equation}

Note that for $v_x=v_y=v_F$,
$\gamma=0$, the results reported for gapped ($\Delta\neq 0$) or ungapped ($\Delta=0$) graphene \cite{gumbs2017}
can be recovered. For $\Delta=0$, the results (\ref{DOS_Verma}) and (\ref{nF}) give the DOS and the doping density of
the tilted system. Table \ref{table1} summarizes the expressions $D(\varepsilon)$ and $n$ when $\gamma\Delta=0$ and $\gamma\Delta\neq 0$.

In Fig.\,\ref{fig1}(a), we show the DOS for several Dirac materials. The gapless cases exhibit the usual linear behavior with energy (green solid and dashed lines). The difference in slope arises from two factors: (i) the factor $(1-\gamma^2)^{-3/2}$ due to the tilting, and (ii) the anisotropy of the velocity. The anisotropy introduces a further rise in the DOS as the velocity $v_F$ of isotropic bands is replaced by the geometric mean $\sqrt{v_xv_y}$. Finally, the DOS displays a step in the massive cases (purple and green dashed lines).
In Fig.\,\ref{fig1}(b), we present the corresponding chemical potentials $\mu(T)$ calculated from Eq.\,(\ref{trascendental}) at a fixed positive value of $\varepsilon_F$. The function $\mu(T)$ remains positive in all cases. In the absence of tilting, the presence of a gap leads to a more rapid decrease of $\mu(T)$ as a function of temperature compared to the simplest case of graphene, due to the vanishing density of states in the region $|\varepsilon|<\Delta$. However, in the case of massive tilting ($\gamma\Delta\neq0$), the rate of decrease of $\mu(T)$ decelerates compared to gapped graphene, as the window of vanishing density of states is reduced to $|\varepsilon|<\tilde{\Delta}$.
This behavior should be contrasted with certain direct-band-gap transition-metal dichalcogenides with broken electron-hole symmetry, such as MoS$_2$, where the chemical potential for electron doping switches from positive to negative at sufficiently high temperatures \cite{gumbs2017}.

In Fig.\,\ref{Fig2} the spectrum of reflectivity $R$ and the corresponding  thermal difference 
$\Delta R/R$ are shown for gapless Dirac systems and several values of temperature at fixed $\Delta T$. Fig.\,\ref{Fig2}(a) shows the isotropic case of doped graphene with a single optical threshold at frequency $2\varepsilon_F$ in the reflectivity spectrum at zero temperature.
As the temperature increases, the spectral signature of the van-Hove singularity is smoothed due to thermal broadening.
However, the modulated spectrum shows a sharp feature around $2\varepsilon_F$ (Fig.\,\ref{Fig2}(b)) still 
discernible for high temperatures. 
When the Dirac cone is tilted, the optical threshold at twice the Fermi energy splits into a couple of critical frequencies $\hbar\omega_{\pm}=2|\varepsilon_F|/(1\mp\gamma)$ 
\cite{verma,seminal} as shown in the spectrum of reflectivity at zero temperature in Fig.\,\ref{Fig2}(c). 
The sharp definition of such features is rapidly lost for increasing temperature. 
Notwithstanding, the thermal difference emphasizes $\hbar\omega_{\pm}$ as expected (Fig.\,\ref{Fig2}(d))

\begin{figure}
    \centering
    \includegraphics[scale=0.59]{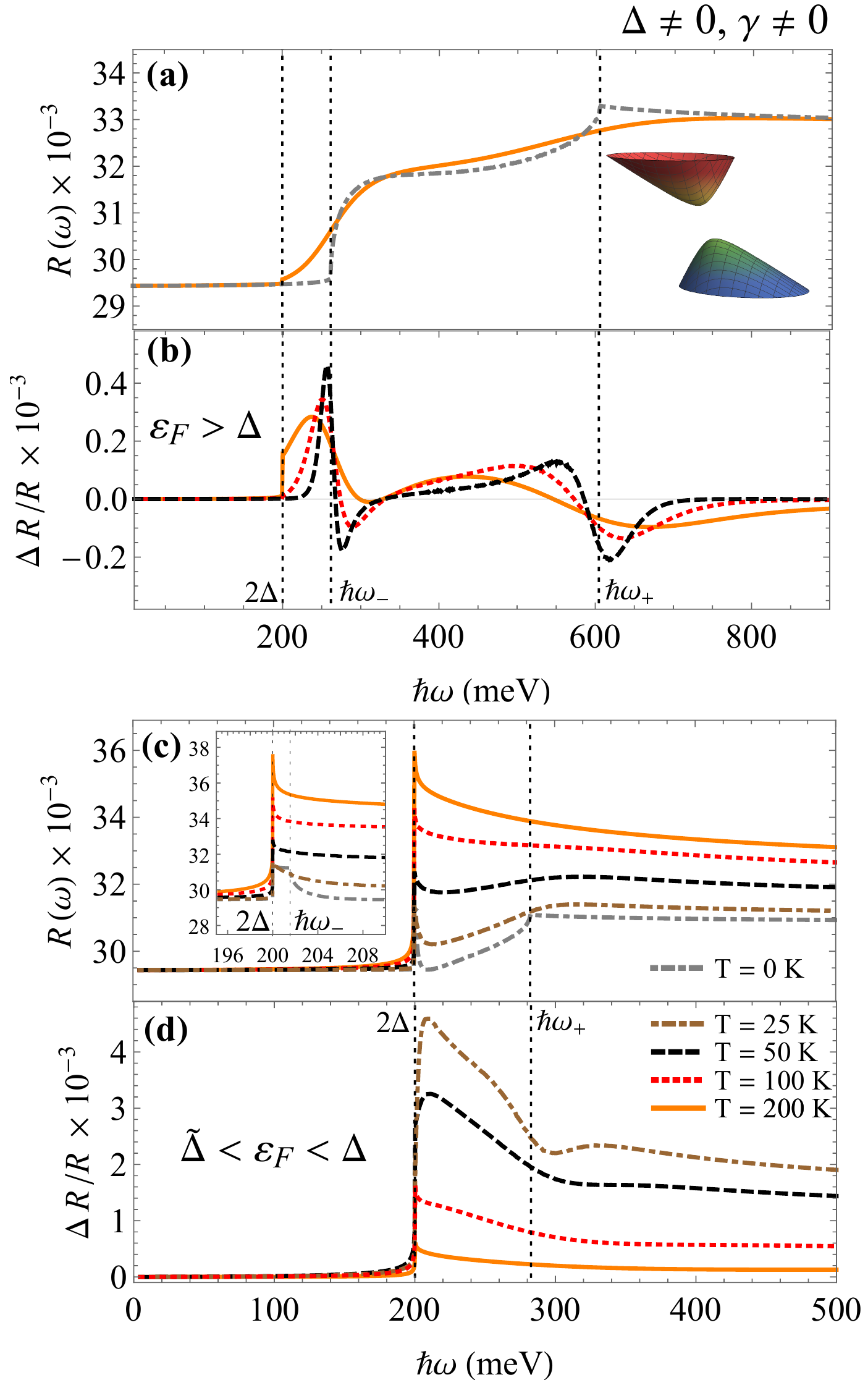}
    \caption{Optical reflectivity and thermal derivative of a massive tilted Dirac system ($\Delta=100$ meV) for several temperatures when the Fermi energy lies (a)-(b) above the gap $\varepsilon_F>\Delta$ (with $\varepsilon_F=170$ meV) and (c)-(d) in the indirect zone
    $\tilde{\Delta}<\varepsilon_F<\Delta$ (with $\varepsilon_F=95$ meV). The velocities $v_i$ are taken as in borophene, and $\epsilon_2=2\epsilon_1=2$, $\Delta T=1\,$K.}
    \label{Fig4}
\end{figure}

The case of a gapped material with cones without tilting is illustrated in 
Fig.\,\ref{Fig3} for gapped graphene. The spectra look similar to those of the gapless graphene (Fig.\,\ref{Fig2}(a)-(b)), but now a feature appears at the energy gap $2\Delta$ due to the onset for interband transitions. If $\varepsilon_F$ lies within the gap, then
this onset is the only salient feature in the spectra.

Figure\,\ref{Fig4} shows the results for a massive tilted system
($\gamma\Delta\neq 0$), corresponding to the cases $\varepsilon_F>\Delta$ (Fig.\,\ref{Fig4}(a)-(b)) and
$\tilde{\Delta}<\varepsilon_F<\Delta$ (Fig.\,\ref{Fig4}(c)-(d)). In the former case, the joint density of states (JDOS) at zero temperature displays two van-Hove singularities at the energies \cite{seminal}
\begin{equation}
\hbar\omega_{\pm}=\frac{2}{1-\gamma^2}\left(|\varepsilon_F|\pm\gamma
\sqrt{\varepsilon_F^2-\tilde{\Delta}^2}\right) \,, 
\end{equation}
and the optical conductivity tensor looks
qualitatively similar to that of the case $\Delta=0, \gamma\neq 0$ (Fig.\,\ref{Fig2}(c))
\cite{verma,seminal}. As in the case of gapped graphene (Fig.\,\ref{Fig3}), the reflectivity spectrum and its thermal difference show an optical feature at 2$\Delta$ due to the onset of interband transitions.
In contrast, in the latter case the JDOS develops three
critical points, at $2\Delta$, $\hbar\omega_-$, and $\hbar\omega_+$, and
a reduced overall size in comparison to the cases $\varepsilon_F<\tilde{\Delta}$
and $\varepsilon_F>\Delta$. Moreover, the number of interband transitions is strongly diminished between $\hbar\omega_-$ and $\hbar\omega_+$ because the $k$-space available 
for transitions is considerably reduced. This behavior and the appearance of three critical points constitute an optical signature of the indirect gap \cite{seminal}. 
As can be seen
in Fig.\,\ref{Fig4}(c), the feature at $\hbar\omega_-$ is lost even at small temperatures, and the derivative spectra do not resolve 
the critical frequencies clearly (Fig.\,\ref{Fig4}(d)). 
Moreover, besides the expected broadening, the reflectivity $R(\omega,T)$ increases slightly with temperature. This is due to the fact that as the temperatures increases
the function $\partial f/\partial\mu'=
[4k_BT\cosh^2(\mu(T)-\mu')]^{-1}$ is sampling the zero-temperature response $\sigma_{xx}(\omega,0,\mu')$ in a $\mu'$-region where it is increasing appreciably \cite{seminal}.
Quantitatively, the integral (\ref{sigmaT}) increases mildly as is shown in 
Fig.\,\ref{Fig4}(c). This is in contrast to the situation observed in Fig.\,\ref{Fig4}(a),
where the convolution process involves a $\mu'$-region where the zero-temperature
$\sigma_{xx}$ does not present significant variation of its magnitude, leading just
to a broadening.
Note however that at finite temperature, the indirect nature of the gap is still traceable from the spectra in the case $\varepsilon_F>\Delta$ 
(Fig.\,\ref{Fig4}(a)-(b)), because the presence of the two features at $\hbar\omega_{\pm}$
suggests tilted cones, while the discontinuity at $2\Delta$ reflects a gapped system.

For dopped system at finite temperature, the knowledge
of the critical frequencies $\omega_{\pm}$ from the thermal difference spectrum suggests a neat way to find the tilting parameter $\gamma$ and the energy gap $\varepsilon_g=2\tilde{\Delta}$. 
Using the definition of the $k$-th power mean,
$M_k=\{[(\hbar\omega_+)^k+(\hbar\omega_-)^k]/2\}^{1/k}$, we find
\begin{eqnarray}
\gamma^2 &=& 1-2|\varepsilon_F|/M_1\,, \\
\varepsilon_g &=& 2|\varepsilon_F|\left(\frac{M_{-1}-2|\varepsilon_F|}{M_1-2|\varepsilon_F|}\right)^{1/2}\,,
\end{eqnarray}
where $M_1=(\hbar\omega_++\hbar\omega_-)/2$ and
$M_{-1}=2(\hbar\omega_+)(\hbar\omega_-)/(\hbar\omega_++\hbar\omega_-)$ are the arithmetic and harmonic means of the numbers
$\hbar\omega_{\pm}$, respectively. 

In summary, we explore the joint density of states of doped massive tilted 2D Dirac 
systems through the temperature dependence of the reflectivity spectrum $R(\omega,T)$. 
The tilt of the bands increase the density of states and decrease the chemical potential
function $\mu(T)$. On the other hand, the
indirect gap of the energy dispersion modifies notably the optical conductivity spectrum as a function of the Fermi energy.
Based on the difference of the reflectivity when the sample is held at two close temperatures, we obtain the fractional change $\Delta R/R$ which emphasizes the spectral structure associated to interband transitions and probe its change due to variation of the Fermi level. The measurement of the critical energies in the optical response would allow to estimate the magnitude of the tilting parameter and energy gap.
Our results suggest that thermal difference spectroscopy is a plausible/possible optical technique to identify such critical points despite the broadening of the
spectra at finite temperature.
The overall size of the shown derivative spectra is of the same order of magnitude
of the measured signals attained with current tools and they
should be experimentally accesible. We hope that this work will
stimulate further experiments and theory.

M.A.M. acknowledges support from U.S. Department of Energy, Office of Basic Energy Sciences, Materials Science and Engineering
Division.
\bibliography{biblio.bib}

\begin{thebibliography}{55}%
\makeatletter
\providecommand \@ifxundefined [1]{%
 \@ifx{#1\undefined}
}%
\providecommand \@ifnum [1]{%
 \ifnum #1\expandafter \@firstoftwo
 \else \expandafter \@secondoftwo
 \fi
}%
\providecommand \@ifx [1]{%
 \ifx #1\expandafter \@firstoftwo
 \else \expandafter \@secondoftwo
 \fi
}%
\providecommand \natexlab [1]{#1}%
\providecommand \enquote  [1]{``#1''}%
\providecommand \bibnamefont  [1]{#1}%
\providecommand \bibfnamefont [1]{#1}%
\providecommand \citenamefont [1]{#1}%
\providecommand \href@noop [0]{\@secondoftwo}%
\providecommand \href [0]{\begingroup \@sanitize@url \@href}%
\providecommand \@href[1]{\@@startlink{#1}\@@href}%
\providecommand \@@href[1]{\endgroup#1\@@endlink}%
\providecommand \@sanitize@url [0]{\catcode `\\12\catcode `\$12\catcode
  `\&12\catcode `\#12\catcode `\^12\catcode `\_12\catcode `\%12\relax}%
\providecommand \@@startlink[1]{}%
\providecommand \@@endlink[0]{}%
\providecommand \url  [0]{\begingroup\@sanitize@url \@url }%
\providecommand \@url [1]{\endgroup\@href {#1}{\urlprefix }}%
\providecommand \urlprefix  [0]{URL }%
\providecommand \Eprint [0]{\href }%
\providecommand \doibase [0]{https://doi.org/}%
\providecommand \selectlanguage [0]{\@gobble}%
\providecommand \bibinfo  [0]{\@secondoftwo}%
\providecommand \bibfield  [0]{\@secondoftwo}%
\providecommand \translation [1]{[#1]}%
\providecommand \BibitemOpen [0]{}%
\providecommand \bibitemStop [0]{}%
\providecommand \bibitemNoStop [0]{.\EOS\space}%
\providecommand \EOS [0]{\spacefactor3000\relax}%
\providecommand \BibitemShut  [1]{\csname bibitem#1\endcsname}%
\let\auto@bib@innerbib\@empty
\bibitem [{\citenamefont {Halevi}(1995)}]{PhotoProbe}%
  \BibitemOpen
  \bibinfo {editor} {\bibfnamefont {P.}~\bibnamefont {Halevi}},\ ed.,\
  \href@noop {} {\emph {\bibinfo {title} {Photonic probes of surfaces}}},\
  Vol.~\bibinfo {volume} {2}\ (\bibinfo  {publisher} {Elsevier Science B.V.},\
  \bibinfo {address} {North-Holland, Amsterdam},\ \bibinfo {year}
  {1995})\BibitemShut {NoStop}%
\bibitem [{\citenamefont {McGilp}(2010)}]{ProbSurf}%
  \BibitemOpen
  \bibfield  {author} {\bibinfo {author} {\bibfnamefont {J.~F.}\ \bibnamefont
  {McGilp}},\ }\bibfield  {title} {\bibinfo {title} {Probing surface and
  interface structure using optics},\ }\href
  {https://doi.org/10.1088/0953-8984/22/8/084018} {\bibfield  {journal}
  {\bibinfo  {journal} {Journal of Physics: Condensed Matter}\ }\textbf
  {\bibinfo {volume} {22}},\ \bibinfo {pages} {084018} (\bibinfo {year}
  {2010})}\BibitemShut {NoStop}%
\bibitem [{\citenamefont {Novotny}\ and\ \citenamefont
  {Hecht}(2012)}]{novotny2012principles}%
  \BibitemOpen
  \bibfield  {author} {\bibinfo {author} {\bibfnamefont {L.}~\bibnamefont
  {Novotny}}\ and\ \bibinfo {author} {\bibfnamefont {B.}~\bibnamefont
  {Hecht}},\ }\href@noop {} {\emph {\bibinfo {title} {Principles of
  nano-optics}}}\ (\bibinfo  {publisher} {Cambridge university press},\
  \bibinfo {year} {2012})\BibitemShut {NoStop}%
\bibitem [{\citenamefont {Downer}(2016)}]{Downer2016}%
  \BibitemOpen
  \bibfield  {author} {\bibinfo {author} {\bibfnamefont {M.}~\bibnamefont
  {Downer}},\ }\bibfield  {title} {\bibinfo {title} {Optics of surfaces and
  interfaces},\ }\href {https://doi.org/https://doi.org/10.1002/pssb.201670514}
  {\bibfield  {journal} {\bibinfo  {journal} {physica status solidi (b)}\
  }\textbf {\bibinfo {volume} {253}},\ \bibinfo {pages} {197} (\bibinfo {year}
  {2016})}\BibitemShut {NoStop}%
\bibitem [{\citenamefont {Zhang}\ \emph {et~al.}(2013)\citenamefont {Zhang},
  \citenamefont {Han}, \citenamefont {Wu}, \citenamefont {Jasensky},\ and\
  \citenamefont {Chen}}]{Zhang2013}%
  \BibitemOpen
  \bibfield  {author} {\bibinfo {author} {\bibfnamefont {X.}~\bibnamefont
  {Zhang}}, \bibinfo {author} {\bibfnamefont {X.}~\bibnamefont {Han}}, \bibinfo
  {author} {\bibfnamefont {F.}~\bibnamefont {Wu}}, \bibinfo {author}
  {\bibfnamefont {J.}~\bibnamefont {Jasensky}},\ and\ \bibinfo {author}
  {\bibfnamefont {Z.}~\bibnamefont {Chen}},\ }\bibfield  {title} {\bibinfo
  {title} {Nano-bio interfaces probed by advanced optical spectroscopy: {F}rom
  model system studies to optical biosensors},\ }\href
  {https://doi.org/10.1007/s11434-013-5700-y} {\bibfield  {journal} {\bibinfo
  {journal} {Chinese Science Bulletin}\ }\textbf {\bibinfo {volume} {58}},\
  \bibinfo {pages} {2537} (\bibinfo {year} {2013})}\BibitemShut {NoStop}%
\bibitem [{\citenamefont {Prévot}\ \emph {et~al.}(2016)\citenamefont
  {Prévot}, \citenamefont {Bernard}, \citenamefont {Cruguel}, \citenamefont
  {Curcella}, \citenamefont {Lazzeri}, \citenamefont {Leoni}, \citenamefont
  {Masson}, \citenamefont {Ranguis},\ and\ \citenamefont
  {Borensztein}}]{Geoffroy2016}%
  \BibitemOpen
  \bibfield  {author} {\bibinfo {author} {\bibfnamefont {G.}~\bibnamefont
  {Prévot}}, \bibinfo {author} {\bibfnamefont {R.}~\bibnamefont {Bernard}},
  \bibinfo {author} {\bibfnamefont {H.}~\bibnamefont {Cruguel}}, \bibinfo
  {author} {\bibfnamefont {A.}~\bibnamefont {Curcella}}, \bibinfo {author}
  {\bibfnamefont {M.}~\bibnamefont {Lazzeri}}, \bibinfo {author} {\bibfnamefont
  {T.}~\bibnamefont {Leoni}}, \bibinfo {author} {\bibfnamefont
  {L.}~\bibnamefont {Masson}}, \bibinfo {author} {\bibfnamefont
  {A.}~\bibnamefont {Ranguis}},\ and\ \bibinfo {author} {\bibfnamefont
  {Y.}~\bibnamefont {Borensztein}},\ }\bibfield  {title} {\bibinfo {title}
  {Formation of silicene on silver: Strong interaction between {A}g and {S}i},\
  }\href {https://doi.org/https://doi.org/10.1002/pssb.201552524} {\bibfield
  {journal} {\bibinfo  {journal} {physica status solidi (b)}\ }\textbf
  {\bibinfo {volume} {253}},\ \bibinfo {pages} {206} (\bibinfo {year}
  {2016})}\BibitemShut {NoStop}%
\bibitem [{\citenamefont {Sarychev}\ \emph {et~al.}(2022)\citenamefont
  {Sarychev}, \citenamefont {Ivanov}, \citenamefont {Lagarkov}, \citenamefont
  {Ryzhikov}, \citenamefont {Afanasev}, \citenamefont {Bykov}, \citenamefont
  {Barbillon}, \citenamefont {Bakholdin}, \citenamefont {Mikhailov},
  \citenamefont {Smyk}, \citenamefont {Shurygin},\ and\ \citenamefont
  {Shalygin}}]{Sarychev2022}%
  \BibitemOpen
  \bibfield  {author} {\bibinfo {author} {\bibfnamefont {A.~K.}\ \bibnamefont
  {Sarychev}}, \bibinfo {author} {\bibfnamefont {A.}~\bibnamefont {Ivanov}},
  \bibinfo {author} {\bibfnamefont {A.~N.}\ \bibnamefont {Lagarkov}}, \bibinfo
  {author} {\bibfnamefont {I.}~\bibnamefont {Ryzhikov}}, \bibinfo {author}
  {\bibfnamefont {K.}~\bibnamefont {Afanasev}}, \bibinfo {author}
  {\bibfnamefont {I.}~\bibnamefont {Bykov}}, \bibinfo {author} {\bibfnamefont
  {G.}~\bibnamefont {Barbillon}}, \bibinfo {author} {\bibfnamefont
  {N.}~\bibnamefont {Bakholdin}}, \bibinfo {author} {\bibfnamefont
  {M.}~\bibnamefont {Mikhailov}}, \bibinfo {author} {\bibfnamefont
  {A.}~\bibnamefont {Smyk}}, \bibinfo {author} {\bibfnamefont {A.}~\bibnamefont
  {Shurygin}},\ and\ \bibinfo {author} {\bibfnamefont {A.}~\bibnamefont
  {Shalygin}},\ }\bibfield  {title} {\bibinfo {title} {Plasmon localization and
  giant fields in an open-resonator metasurface
  forsurface-enhanced-raman-scattering sensors},\ }\href
  {https://doi.org/10.1103/PhysRevApplied.17.044029} {\bibfield  {journal}
  {\bibinfo  {journal} {Phys. Rev. Appl.}\ }\textbf {\bibinfo {volume} {17}},\
  \bibinfo {pages} {044029} (\bibinfo {year} {2022})}\BibitemShut {NoStop}%
\bibitem [{\citenamefont {Chang}\ \emph {et~al.}(2018)\citenamefont {Chang},
  \citenamefont {Guo},\ and\ \citenamefont {Ni}}]{Chang2018}%
  \BibitemOpen
  \bibfield  {author} {\bibinfo {author} {\bibfnamefont {S.}~\bibnamefont
  {Chang}}, \bibinfo {author} {\bibfnamefont {X.}~\bibnamefont {Guo}},\ and\
  \bibinfo {author} {\bibfnamefont {X.}~\bibnamefont {Ni}},\ }\bibfield
  {title} {\bibinfo {title} {Optical metasurfaces: Progress and applications},\
  }\href {https://doi.org/10.1146/annurev-matsci-070616-124220} {\bibfield
  {journal} {\bibinfo  {journal} {Annual Review of Materials Research}\
  }\textbf {\bibinfo {volume} {48}},\ \bibinfo {pages} {279} (\bibinfo {year}
  {2018})}\BibitemShut {NoStop}%
\bibitem [{\citenamefont {Huang}\ \emph {et~al.}(2023)\citenamefont {Huang},
  \citenamefont {Wang}, \citenamefont {Xie}, \citenamefont {Yu},\ and\
  \citenamefont {Yan}}]{HuangOpt2D}%
  \BibitemOpen
  \bibfield  {author} {\bibinfo {author} {\bibfnamefont {S.}~\bibnamefont
  {Huang}}, \bibinfo {author} {\bibfnamefont {C.}~\bibnamefont {Wang}},
  \bibinfo {author} {\bibfnamefont {Y.}~\bibnamefont {Xie}}, \bibinfo {author}
  {\bibfnamefont {B.}~\bibnamefont {Yu}},\ and\ \bibinfo {author}
  {\bibfnamefont {H.}~\bibnamefont {Yan}},\ }\bibfield  {title} {\bibinfo
  {title} {Optical properties and polaritons of low symmetry 2{D} materials},\
  }\href {https://www.researching.cn/articles/OJc7213c8b23cb9c64} {\bibfield
  {journal} {\bibinfo  {journal} {Photonics Insights}\ }\textbf {\bibinfo
  {volume} {2}},\ \bibinfo {pages} {R03} (\bibinfo {year} {2023})}\BibitemShut
  {NoStop}%
\bibitem [{\citenamefont {Ma}\ \emph {et~al.}(2021)\citenamefont {Ma},
  \citenamefont {Ren}, \citenamefont {Xu},\ and\ \citenamefont
  {Ou}}]{MaTunable}%
  \BibitemOpen
  \bibfield  {author} {\bibinfo {author} {\bibfnamefont {Q.}~\bibnamefont
  {Ma}}, \bibinfo {author} {\bibfnamefont {G.}~\bibnamefont {Ren}}, \bibinfo
  {author} {\bibfnamefont {K.}~\bibnamefont {Xu}},\ and\ \bibinfo {author}
  {\bibfnamefont {J.~Z.}\ \bibnamefont {Ou}},\ }\bibfield  {title} {\bibinfo
  {title} {Tunable optical properties of 2{D} materials and their
  applications},\ }\href
  {https://doi.org/https://doi.org/10.1002/adom.202001313} {\bibfield
  {journal} {\bibinfo  {journal} {Advanced Optical Materials}\ }\textbf
  {\bibinfo {volume} {9}},\ \bibinfo {pages} {2001313} (\bibinfo {year}
  {2021})}\BibitemShut {NoStop}%
\bibitem [{\citenamefont {Gan}\ \emph {et~al.}()\citenamefont {Gan},
  \citenamefont {Englund}, \citenamefont {Van~Thourhout},\ and\ \citenamefont
  {Zhao}}]{Gan2Dmat}%
  \BibitemOpen
  \bibfield  {author} {\bibinfo {author} {\bibfnamefont {X.}~\bibnamefont
  {Gan}}, \bibinfo {author} {\bibfnamefont {D.}~\bibnamefont {Englund}},
  \bibinfo {author} {\bibfnamefont {D.}~\bibnamefont {Van~Thourhout}},\ and\
  \bibinfo {author} {\bibfnamefont {J.}~\bibnamefont {Zhao}},\ }\bibfield
  {title} {\bibinfo {title} {2{D} materials-enabled optical modulators: {F}rom
  visible to terahertz spectral range},\ }\bibfield  {journal} {\bibinfo
  {journal} {Applied Physics Reviews}\ }\textbf {\bibinfo {volume} {9}},\ \href
  {https://doi.org/10.1063/5.0078416} {10.1063/5.0078416},\ \bibinfo {note}
  {021302}\BibitemShut {NoStop}%
\bibitem [{\citenamefont {Xin}\ \emph {et~al.}(2019)\citenamefont {Xin},
  \citenamefont {Jiang}, \citenamefont {Sun}, \citenamefont {Gao},
  \citenamefont {Chen}, \citenamefont {Zhao}, \citenamefont {Yang},
  \citenamefont {Liu}, \citenamefont {Tian},\ and\ \citenamefont
  {Guo}}]{XinBP}%
  \BibitemOpen
  \bibfield  {author} {\bibinfo {author} {\bibfnamefont {W.}~\bibnamefont
  {Xin}}, \bibinfo {author} {\bibfnamefont {H.~B.}\ \bibnamefont {Jiang}},
  \bibinfo {author} {\bibfnamefont {T.~Q.}\ \bibnamefont {Sun}}, \bibinfo
  {author} {\bibfnamefont {X.~G.}\ \bibnamefont {Gao}}, \bibinfo {author}
  {\bibfnamefont {S.~N.}\ \bibnamefont {Chen}}, \bibinfo {author}
  {\bibfnamefont {B.}~\bibnamefont {Zhao}}, \bibinfo {author} {\bibfnamefont
  {J.~J.}\ \bibnamefont {Yang}}, \bibinfo {author} {\bibfnamefont {Z.~B.}\
  \bibnamefont {Liu}}, \bibinfo {author} {\bibfnamefont {J.~G.}\ \bibnamefont
  {Tian}},\ and\ \bibinfo {author} {\bibfnamefont {C.~L.}\ \bibnamefont
  {Guo}},\ }\bibfield  {title} {\bibinfo {title} {Optical anisotropy of black
  phosphorus by total internal reflection},\ }\href
  {https://doi.org/https://doi.org/10.1016/j.nanoms.2019.09.006} {\bibfield
  {journal} {\bibinfo  {journal} {Nano Materials Science}\ }\textbf {\bibinfo
  {volume} {1}},\ \bibinfo {pages} {304} (\bibinfo {year} {2019})}\BibitemShut
  {NoStop}%
\bibitem [{\citenamefont {Wang}\ \emph {et~al.}(2020)\citenamefont {Wang},
  \citenamefont {Zhang}, \citenamefont {Huang}, \citenamefont {Xie},\ and\
  \citenamefont {Yan}}]{WangBP}%
  \BibitemOpen
  \bibfield  {author} {\bibinfo {author} {\bibfnamefont {C.}~\bibnamefont
  {Wang}}, \bibinfo {author} {\bibfnamefont {G.}~\bibnamefont {Zhang}},
  \bibinfo {author} {\bibfnamefont {S.}~\bibnamefont {Huang}}, \bibinfo
  {author} {\bibfnamefont {Y.}~\bibnamefont {Xie}},\ and\ \bibinfo {author}
  {\bibfnamefont {H.}~\bibnamefont {Yan}},\ }\bibfield  {title} {\bibinfo
  {title} {The optical properties and plasmonics of anisotropic 2{D}
  materials},\ }\href {https://doi.org/https://doi.org/10.1002/adom.201900996}
  {\bibfield  {journal} {\bibinfo  {journal} {Advanced Optical Materials}\
  }\textbf {\bibinfo {volume} {8}},\ \bibinfo {pages} {1900996} (\bibinfo
  {year} {2020})}\BibitemShut {NoStop}%
\bibitem [{\citenamefont {Zhang}\ \emph {et~al.}(2019)\citenamefont {Zhang},
  \citenamefont {Zhang}, \citenamefont {Chen}, \citenamefont {Shen},
  \citenamefont {An}, \citenamefont {Hu}, \citenamefont {Dong}, \citenamefont
  {Liu},\ and\ \citenamefont {Zhu}}]{ZhangWTe2}%
  \BibitemOpen
  \bibfield  {author} {\bibinfo {author} {\bibfnamefont {Q.}~\bibnamefont
  {Zhang}}, \bibinfo {author} {\bibfnamefont {R.}~\bibnamefont {Zhang}},
  \bibinfo {author} {\bibfnamefont {J.}~\bibnamefont {Chen}}, \bibinfo {author}
  {\bibfnamefont {W.}~\bibnamefont {Shen}}, \bibinfo {author} {\bibfnamefont
  {C.}~\bibnamefont {An}}, \bibinfo {author} {\bibfnamefont {X.}~\bibnamefont
  {Hu}}, \bibinfo {author} {\bibfnamefont {M.}~\bibnamefont {Dong}}, \bibinfo
  {author} {\bibfnamefont {J.}~\bibnamefont {Liu}},\ and\ \bibinfo {author}
  {\bibfnamefont {L.}~\bibnamefont {Zhu}},\ }\bibfield  {title} {\bibinfo
  {title} {Remarkable electronic and optical anisotropy of layered
  {1T$'$-WTe$_2$} 2{D} materials},\ }\href
  {https://doi.org/10.3762/bjnano.10.170} {\bibfield  {journal} {\bibinfo
  {journal} {Beilstein Journal of Nanotechnology}\ }\textbf {\bibinfo {volume}
  {10}},\ \bibinfo {pages} {1745} (\bibinfo {year} {2019})}\BibitemShut
  {NoStop}%
\bibitem [{\citenamefont {Shen}\ \emph
  {et~al.}(2018{\natexlab{a}})\citenamefont {Shen}, \citenamefont {Hu},
  \citenamefont {Tao}, \citenamefont {Liu}, \citenamefont {Fan}, \citenamefont
  {Wei}, \citenamefont {An}, \citenamefont {Chen}, \citenamefont {Wu},
  \citenamefont {Li}, \citenamefont {Liu}, \citenamefont {Zhang}, \citenamefont
  {Sun},\ and\ \citenamefont {Hu}}]{ShenReS2}%
  \BibitemOpen
  \bibfield  {author} {\bibinfo {author} {\bibfnamefont {W.}~\bibnamefont
  {Shen}}, \bibinfo {author} {\bibfnamefont {C.}~\bibnamefont {Hu}}, \bibinfo
  {author} {\bibfnamefont {J.}~\bibnamefont {Tao}}, \bibinfo {author}
  {\bibfnamefont {J.}~\bibnamefont {Liu}}, \bibinfo {author} {\bibfnamefont
  {S.}~\bibnamefont {Fan}}, \bibinfo {author} {\bibfnamefont {Y.}~\bibnamefont
  {Wei}}, \bibinfo {author} {\bibfnamefont {C.}~\bibnamefont {An}}, \bibinfo
  {author} {\bibfnamefont {J.}~\bibnamefont {Chen}}, \bibinfo {author}
  {\bibfnamefont {S.}~\bibnamefont {Wu}}, \bibinfo {author} {\bibfnamefont
  {Y.}~\bibnamefont {Li}}, \bibinfo {author} {\bibfnamefont {J.}~\bibnamefont
  {Liu}}, \bibinfo {author} {\bibfnamefont {D.}~\bibnamefont {Zhang}}, \bibinfo
  {author} {\bibfnamefont {L.}~\bibnamefont {Sun}},\ and\ \bibinfo {author}
  {\bibfnamefont {X.}~\bibnamefont {Hu}},\ }\bibfield  {title} {\bibinfo
  {title} {Resolving the optical anisotropy of low-symmetry 2{D} materials},\
  }\href {https://doi.org/10.1039/C7NR09173G} {\bibfield  {journal} {\bibinfo
  {journal} {Nanoscale}\ }\textbf {\bibinfo {volume} {10}},\ \bibinfo {pages}
  {8329} (\bibinfo {year} {2018}{\natexlab{a}})}\BibitemShut {NoStop}%
\bibitem [{\citenamefont {Li}\ \emph {et~al.}(2019)\citenamefont {Li},
  \citenamefont {Han}, \citenamefont {Pi}, \citenamefont {Niu}, \citenamefont
  {Han}, \citenamefont {Wang}, \citenamefont {Su}, \citenamefont {Li},
  \citenamefont {Xiong}, \citenamefont {Bando},\ and\ \citenamefont
  {Zhai}}]{LiEmerg}%
  \BibitemOpen
  \bibfield  {author} {\bibinfo {author} {\bibfnamefont {L.}~\bibnamefont
  {Li}}, \bibinfo {author} {\bibfnamefont {W.}~\bibnamefont {Han}}, \bibinfo
  {author} {\bibfnamefont {L.}~\bibnamefont {Pi}}, \bibinfo {author}
  {\bibfnamefont {P.}~\bibnamefont {Niu}}, \bibinfo {author} {\bibfnamefont
  {J.}~\bibnamefont {Han}}, \bibinfo {author} {\bibfnamefont {C.}~\bibnamefont
  {Wang}}, \bibinfo {author} {\bibfnamefont {B.}~\bibnamefont {Su}}, \bibinfo
  {author} {\bibfnamefont {H.}~\bibnamefont {Li}}, \bibinfo {author}
  {\bibfnamefont {J.}~\bibnamefont {Xiong}}, \bibinfo {author} {\bibfnamefont
  {Y.}~\bibnamefont {Bando}},\ and\ \bibinfo {author} {\bibfnamefont
  {T.}~\bibnamefont {Zhai}},\ }\bibfield  {title} {\bibinfo {title} {Emerging
  in-plane anisotropic two-dimensional materials},\ }\href
  {https://doi.org/https://doi.org/10.1002/inf2.12005} {\bibfield  {journal}
  {\bibinfo  {journal} {InfoMat}\ }\textbf {\bibinfo {volume} {1}},\ \bibinfo
  {pages} {54} (\bibinfo {year} {2019})}\BibitemShut {NoStop}%
\bibitem [{\citenamefont {Frisenda}\ \emph {et~al.}(2017)\citenamefont
  {Frisenda}, \citenamefont {Niu}, \citenamefont {Gant}, \citenamefont
  {Molina-Mendoza}, \citenamefont {Schmidt}, \citenamefont {Bratschitsch},
  \citenamefont {Liu}, \citenamefont {Fu}, \citenamefont {Dumcenco},
  \citenamefont {Kis}, \citenamefont {Lara},\ and\ \citenamefont
  {Castellanos-Gomez}}]{FrisendaMicro}%
  \BibitemOpen
  \bibfield  {author} {\bibinfo {author} {\bibfnamefont {R.}~\bibnamefont
  {Frisenda}}, \bibinfo {author} {\bibfnamefont {Y.}~\bibnamefont {Niu}},
  \bibinfo {author} {\bibfnamefont {P.}~\bibnamefont {Gant}}, \bibinfo {author}
  {\bibfnamefont {A.~J.}\ \bibnamefont {Molina-Mendoza}}, \bibinfo {author}
  {\bibfnamefont {R.}~\bibnamefont {Schmidt}}, \bibinfo {author} {\bibfnamefont
  {R.}~\bibnamefont {Bratschitsch}}, \bibinfo {author} {\bibfnamefont
  {J.}~\bibnamefont {Liu}}, \bibinfo {author} {\bibfnamefont {L.}~\bibnamefont
  {Fu}}, \bibinfo {author} {\bibfnamefont {D.}~\bibnamefont {Dumcenco}},
  \bibinfo {author} {\bibfnamefont {A.}~\bibnamefont {Kis}}, \bibinfo {author}
  {\bibfnamefont {D.~P.~D.}\ \bibnamefont {Lara}},\ and\ \bibinfo {author}
  {\bibfnamefont {A.}~\bibnamefont {Castellanos-Gomez}},\ }\bibfield  {title}
  {\bibinfo {title} {Micro-reflectance and transmittance spectroscopy: a
  versatile and powerful tool to characterize 2{D} materials},\ }\href
  {https://doi.org/10.1088/1361-6463/aa5256} {\bibfield  {journal} {\bibinfo
  {journal} {Journal of Physics D: Applied Physics}\ }\textbf {\bibinfo
  {volume} {50}},\ \bibinfo {pages} {074002} (\bibinfo {year}
  {2017})}\BibitemShut {NoStop}%
\bibitem [{\citenamefont {Zhou}\ \emph {et~al.}(2019)\citenamefont {Zhou},
  \citenamefont {Cui}, \citenamefont {Tan}, \citenamefont {Liu},\ and\
  \citenamefont {Wei}}]{ZhouOpt}%
  \BibitemOpen
  \bibfield  {author} {\bibinfo {author} {\bibfnamefont {Z.}~\bibnamefont
  {Zhou}}, \bibinfo {author} {\bibfnamefont {Y.}~\bibnamefont {Cui}}, \bibinfo
  {author} {\bibfnamefont {P.-H.}\ \bibnamefont {Tan}}, \bibinfo {author}
  {\bibfnamefont {X.}~\bibnamefont {Liu}},\ and\ \bibinfo {author}
  {\bibfnamefont {Z.}~\bibnamefont {Wei}},\ }\bibfield  {title} {\bibinfo
  {title} {Optical and electrical properties of two-dimensional anisotropic
  materials},\ }\href {https://doi.org/10.1088/1674-4926/40/6/061001}
  {\bibfield  {journal} {\bibinfo  {journal} {Journal of Semiconductors}\
  }\textbf {\bibinfo {volume} {40}},\ \bibinfo {pages} {061001} (\bibinfo
  {year} {2019})}\BibitemShut {NoStop}%
\bibitem [{\citenamefont {Sun}\ \emph {et~al.}(2017)\citenamefont {Sun},
  \citenamefont {Gu}, \citenamefont {Zeng}, \citenamefont {Huang},
  \citenamefont {Yan}, \citenamefont {Liu}, \citenamefont {Yang},\ and\
  \citenamefont {Koh}}]{SunTemp}%
  \BibitemOpen
  \bibfield  {author} {\bibinfo {author} {\bibfnamefont {B.}~\bibnamefont
  {Sun}}, \bibinfo {author} {\bibfnamefont {X.}~\bibnamefont {Gu}}, \bibinfo
  {author} {\bibfnamefont {Q.}~\bibnamefont {Zeng}}, \bibinfo {author}
  {\bibfnamefont {X.}~\bibnamefont {Huang}}, \bibinfo {author} {\bibfnamefont
  {Y.}~\bibnamefont {Yan}}, \bibinfo {author} {\bibfnamefont {Z.}~\bibnamefont
  {Liu}}, \bibinfo {author} {\bibfnamefont {R.}~\bibnamefont {Yang}},\ and\
  \bibinfo {author} {\bibfnamefont {Y.~K.}\ \bibnamefont {Koh}},\ }\bibfield
  {title} {\bibinfo {title} {Temperature dependence of anisotropic
  thermal-conductivity tensor of bulk black phosphorus},\ }\href
  {https://doi.org/https://doi.org/10.1002/adma.201603297} {\bibfield
  {journal} {\bibinfo  {journal} {Advanced Materials}\ }\textbf {\bibinfo
  {volume} {29}},\ \bibinfo {pages} {1603297} (\bibinfo {year}
  {2017})}\BibitemShut {NoStop}%
\bibitem [{\citenamefont {Shen}\ \emph
  {et~al.}(2018{\natexlab{b}})\citenamefont {Shen}, \citenamefont {Hu},
  \citenamefont {Tao}, \citenamefont {Liu}, \citenamefont {Fan}, \citenamefont
  {Wei}, \citenamefont {An}, \citenamefont {Chen}, \citenamefont {Wu},
  \citenamefont {Li}, \citenamefont {Liu}, \citenamefont {Zhang}, \citenamefont
  {Sun},\ and\ \citenamefont {Hu}}]{ShenResolv}%
  \BibitemOpen
  \bibfield  {author} {\bibinfo {author} {\bibfnamefont {W.}~\bibnamefont
  {Shen}}, \bibinfo {author} {\bibfnamefont {C.}~\bibnamefont {Hu}}, \bibinfo
  {author} {\bibfnamefont {J.}~\bibnamefont {Tao}}, \bibinfo {author}
  {\bibfnamefont {J.}~\bibnamefont {Liu}}, \bibinfo {author} {\bibfnamefont
  {S.}~\bibnamefont {Fan}}, \bibinfo {author} {\bibfnamefont {Y.}~\bibnamefont
  {Wei}}, \bibinfo {author} {\bibfnamefont {C.}~\bibnamefont {An}}, \bibinfo
  {author} {\bibfnamefont {J.}~\bibnamefont {Chen}}, \bibinfo {author}
  {\bibfnamefont {S.}~\bibnamefont {Wu}}, \bibinfo {author} {\bibfnamefont
  {Y.}~\bibnamefont {Li}}, \bibinfo {author} {\bibfnamefont {J.}~\bibnamefont
  {Liu}}, \bibinfo {author} {\bibfnamefont {D.}~\bibnamefont {Zhang}}, \bibinfo
  {author} {\bibfnamefont {L.}~\bibnamefont {Sun}},\ and\ \bibinfo {author}
  {\bibfnamefont {X.}~\bibnamefont {Hu}},\ }\bibfield  {title} {\bibinfo
  {title} {Resolving the optical anisotropy of low-symmetry 2{D} materials},\
  }\href {https://doi.org/10.1039/C7NR09173G} {\bibfield  {journal} {\bibinfo
  {journal} {Nanoscale}\ }\textbf {\bibinfo {volume} {10}},\ \bibinfo {pages}
  {8329} (\bibinfo {year} {2018}{\natexlab{b}})}\BibitemShut {NoStop}%
\bibitem [{\citenamefont {Tao}\ \emph {et~al.}(2015)\citenamefont {Tao},
  \citenamefont {Shen}, \citenamefont {Wu}, \citenamefont {Liu}, \citenamefont
  {Feng}, \citenamefont {Wang}, \citenamefont {Hu}, \citenamefont {Yao},
  \citenamefont {Zhang}, \citenamefont {Pang}, \citenamefont {Duan},
  \citenamefont {Liu}, \citenamefont {Zhou},\ and\ \citenamefont
  {Zhang}}]{TaoMech}%
  \BibitemOpen
  \bibfield  {author} {\bibinfo {author} {\bibfnamefont {J.}~\bibnamefont
  {Tao}}, \bibinfo {author} {\bibfnamefont {W.}~\bibnamefont {Shen}}, \bibinfo
  {author} {\bibfnamefont {S.}~\bibnamefont {Wu}}, \bibinfo {author}
  {\bibfnamefont {L.}~\bibnamefont {Liu}}, \bibinfo {author} {\bibfnamefont
  {Z.}~\bibnamefont {Feng}}, \bibinfo {author} {\bibfnamefont {C.}~\bibnamefont
  {Wang}}, \bibinfo {author} {\bibfnamefont {C.}~\bibnamefont {Hu}}, \bibinfo
  {author} {\bibfnamefont {P.}~\bibnamefont {Yao}}, \bibinfo {author}
  {\bibfnamefont {H.}~\bibnamefont {Zhang}}, \bibinfo {author} {\bibfnamefont
  {W.}~\bibnamefont {Pang}}, \bibinfo {author} {\bibfnamefont {X.}~\bibnamefont
  {Duan}}, \bibinfo {author} {\bibfnamefont {J.}~\bibnamefont {Liu}}, \bibinfo
  {author} {\bibfnamefont {C.}~\bibnamefont {Zhou}},\ and\ \bibinfo {author}
  {\bibfnamefont {D.}~\bibnamefont {Zhang}},\ }\bibfield  {title} {\bibinfo
  {title} {Mechanical and electrical anisotropy of few-layer black
  phosphorus},\ }\href {https://doi.org/10.1021/acsnano.5b05151} {\bibfield
  {journal} {\bibinfo  {journal} {ACS Nano}\ }\textbf {\bibinfo {volume} {9}},\
  \bibinfo {pages} {11362} (\bibinfo {year} {2015})}\BibitemShut {NoStop}%
\bibitem [{\citenamefont {Verma}\ \emph {et~al.}(2017)\citenamefont {Verma},
  \citenamefont {Mawrie},\ and\ \citenamefont {Ghosh}}]{verma}%
  \BibitemOpen
  \bibfield  {author} {\bibinfo {author} {\bibfnamefont {S.}~\bibnamefont
  {Verma}}, \bibinfo {author} {\bibfnamefont {A.}~\bibnamefont {Mawrie}},\ and\
  \bibinfo {author} {\bibfnamefont {T.~K.}\ \bibnamefont {Ghosh}},\ }\bibfield
  {title} {\bibinfo {title} {Effect of electron-hole asymmetry on optical
  conductivity in {$8\ensuremath{-}Pmmn$} borophene},\ }\href
  {https://doi.org/10.1103/PhysRevB.96.155418} {\bibfield  {journal} {\bibinfo
  {journal} {Phys. Rev. B}\ }\textbf {\bibinfo {volume} {96}},\ \bibinfo
  {pages} {155418} (\bibinfo {year} {2017})}\BibitemShut {NoStop}%
\bibitem [{\citenamefont {Mojarro}\ \emph {et~al.}(2021)\citenamefont
  {Mojarro}, \citenamefont {Carrillo-Bastos},\ and\ \citenamefont
  {Maytorena}}]{seminal}%
  \BibitemOpen
  \bibfield  {author} {\bibinfo {author} {\bibfnamefont {M.~A.}\ \bibnamefont
  {Mojarro}}, \bibinfo {author} {\bibfnamefont {R.}~\bibnamefont
  {Carrillo-Bastos}},\ and\ \bibinfo {author} {\bibfnamefont {J.~A.}\
  \bibnamefont {Maytorena}},\ }\bibfield  {title} {\bibinfo {title} {Optical
  properties of massive anisotropic tilted {D}irac systems},\ }\href
  {https://doi.org/10.1103/PhysRevB.103.165415} {\bibfield  {journal} {\bibinfo
   {journal} {Phys. Rev. B}\ }\textbf {\bibinfo {volume} {103}},\ \bibinfo
  {pages} {165415} (\bibinfo {year} {2021})}\BibitemShut {NoStop}%
\bibitem [{\citenamefont {Mojarro}\ \emph {et~al.}(2022)\citenamefont
  {Mojarro}, \citenamefont {Carrillo-Bastos},\ and\ \citenamefont
  {Maytorena}}]{plasmonsletter}%
  \BibitemOpen
  \bibfield  {author} {\bibinfo {author} {\bibfnamefont {M.~A.}\ \bibnamefont
  {Mojarro}}, \bibinfo {author} {\bibfnamefont {R.}~\bibnamefont
  {Carrillo-Bastos}},\ and\ \bibinfo {author} {\bibfnamefont {J.~A.}\
  \bibnamefont {Maytorena}},\ }\bibfield  {title} {\bibinfo {title} {Hyperbolic
  plasmons in massive tilted two-dimensional {D}irac materials},\ }\href
  {https://doi.org/10.1103/PhysRevB.105.L201408} {\bibfield  {journal}
  {\bibinfo  {journal} {Phys. Rev. B}\ }\textbf {\bibinfo {volume} {105}},\
  \bibinfo {pages} {L201408} (\bibinfo {year} {2022})}\BibitemShut {NoStop}%
\bibitem [{\citenamefont {Tan}\ \emph {et~al.}(2022)\citenamefont {Tan},
  \citenamefont {Hou}, \citenamefont {Yan}, \citenamefont {Guo},\ and\
  \citenamefont {Chang}}]{TanLifs}%
  \BibitemOpen
  \bibfield  {author} {\bibinfo {author} {\bibfnamefont {C.-Y.}\ \bibnamefont
  {Tan}}, \bibinfo {author} {\bibfnamefont {J.-T.}\ \bibnamefont {Hou}},
  \bibinfo {author} {\bibfnamefont {C.-X.}\ \bibnamefont {Yan}}, \bibinfo
  {author} {\bibfnamefont {H.}~\bibnamefont {Guo}},\ and\ \bibinfo {author}
  {\bibfnamefont {H.-R.}\ \bibnamefont {Chang}},\ }\bibfield  {title} {\bibinfo
  {title} {Signatures of {L}ifshitz transition in the optical conductivity of
  two-dimensional tilted {D}irac materials},\ }\href
  {https://doi.org/10.1103/PhysRevB.106.165404} {\bibfield  {journal} {\bibinfo
   {journal} {Phys. Rev. B}\ }\textbf {\bibinfo {volume} {106}},\ \bibinfo
  {pages} {165404} (\bibinfo {year} {2022})}\BibitemShut {NoStop}%
\bibitem [{\citenamefont {Wild}\ \emph {et~al.}(2022)\citenamefont {Wild},
  \citenamefont {Mariani},\ and\ \citenamefont {Portnoi}}]{WildOptical}%
  \BibitemOpen
  \bibfield  {author} {\bibinfo {author} {\bibfnamefont {A.}~\bibnamefont
  {Wild}}, \bibinfo {author} {\bibfnamefont {E.}~\bibnamefont {Mariani}},\ and\
  \bibinfo {author} {\bibfnamefont {M.~E.}\ \bibnamefont {Portnoi}},\
  }\bibfield  {title} {\bibinfo {title} {Optical absorption in two-dimensional
  materials with tilted {D}irac cones},\ }\href
  {https://doi.org/10.1103/PhysRevB.105.205306} {\bibfield  {journal} {\bibinfo
   {journal} {Phys. Rev. B}\ }\textbf {\bibinfo {volume} {105}},\ \bibinfo
  {pages} {205306} (\bibinfo {year} {2022})}\BibitemShut {NoStop}%
\bibitem [{\citenamefont {Tan}\ \emph {et~al.}(2021)\citenamefont {Tan},
  \citenamefont {Yan}, \citenamefont {Zhao}, \citenamefont {Guo},\ and\
  \citenamefont {Chang}}]{TanAnisot}%
  \BibitemOpen
  \bibfield  {author} {\bibinfo {author} {\bibfnamefont {C.-Y.}\ \bibnamefont
  {Tan}}, \bibinfo {author} {\bibfnamefont {C.-X.}\ \bibnamefont {Yan}},
  \bibinfo {author} {\bibfnamefont {Y.-H.}\ \bibnamefont {Zhao}}, \bibinfo
  {author} {\bibfnamefont {H.}~\bibnamefont {Guo}},\ and\ \bibinfo {author}
  {\bibfnamefont {H.-R.}\ \bibnamefont {Chang}},\ }\bibfield  {title} {\bibinfo
  {title} {Anisotropic longitudinal optical conductivities of tilted {D}irac
  bands in
  {$1{T}^{\ensuremath{'}}\text{\ensuremath{-}}\mathrm{Mo}{\mathrm{S}}_{2}$}},\
  }\href {https://doi.org/10.1103/PhysRevB.103.125425} {\bibfield  {journal}
  {\bibinfo  {journal} {Phys. Rev. B}\ }\textbf {\bibinfo {volume} {103}},\
  \bibinfo {pages} {125425} (\bibinfo {year} {2021})}\BibitemShut {NoStop}%
\bibitem [{\citenamefont {Balassis}\ \emph {et~al.}(2022)\citenamefont
  {Balassis}, \citenamefont {Gumbs},\ and\ \citenamefont
  {Roslyak}}]{AntoniosPolariz}%
  \BibitemOpen
  \bibfield  {author} {\bibinfo {author} {\bibfnamefont {A.}~\bibnamefont
  {Balassis}}, \bibinfo {author} {\bibfnamefont {G.}~\bibnamefont {Gumbs}},\
  and\ \bibinfo {author} {\bibfnamefont {O.}~\bibnamefont {Roslyak}},\
  }\bibfield  {title} {\bibinfo {title} {Polarizability, plasmons, and
  screening in {1T$'$-MoS$_2$} with tilted {D}irac bands},\ }\href
  {https://doi.org/https://doi.org/10.1016/j.physleta.2022.128353} {\bibfield
  {journal} {\bibinfo  {journal} {Physics Letters A}\ }\textbf {\bibinfo
  {volume} {449}},\ \bibinfo {pages} {128353} (\bibinfo {year}
  {2022})}\BibitemShut {NoStop}%
\bibitem [{\citenamefont {Fu}\ \emph {et~al.}(2023)\citenamefont {Fu},
  \citenamefont {Zhang}, \citenamefont {Fan}, \citenamefont {Li}, \citenamefont
  {Ma},\ and\ \citenamefont {Liu}}]{Fu2023}%
  \BibitemOpen
  \bibfield  {author} {\bibinfo {author} {\bibfnamefont {B.}~\bibnamefont
  {Fu}}, \bibinfo {author} {\bibfnamefont {R.-W.}\ \bibnamefont {Zhang}},
  \bibinfo {author} {\bibfnamefont {X.}~\bibnamefont {Fan}}, \bibinfo {author}
  {\bibfnamefont {S.}~\bibnamefont {Li}}, \bibinfo {author} {\bibfnamefont
  {D.-S.}\ \bibnamefont {Ma}},\ and\ \bibinfo {author} {\bibfnamefont {C.-C.}\
  \bibnamefont {Liu}},\ }\bibfield  {title} {\bibinfo {title} {2{D} ladder
  polyborane: An ideal {D}irac semimetal with a multi-field-tunable band gap},\
  }\href {https://doi.org/10.1021/acsnano.2c11612} {\bibfield  {journal}
  {\bibinfo  {journal} {ACS Nano}\ }\textbf {\bibinfo {volume} {17}},\ \bibinfo
  {pages} {1638} (\bibinfo {year} {2023})}\BibitemShut {NoStop}%
\bibitem [{\citenamefont {Park}\ \emph {et~al.}(2022)\citenamefont {Park},
  \citenamefont {Sammon}, \citenamefont {Mele},\ and\ \citenamefont
  {Low}}]{Park2022}%
  \BibitemOpen
  \bibfield  {author} {\bibinfo {author} {\bibfnamefont {S.~H.}\ \bibnamefont
  {Park}}, \bibinfo {author} {\bibfnamefont {M.}~\bibnamefont {Sammon}},
  \bibinfo {author} {\bibfnamefont {E.}~\bibnamefont {Mele}},\ and\ \bibinfo
  {author} {\bibfnamefont {T.}~\bibnamefont {Low}},\ }\bibfield  {title}
  {\bibinfo {title} {Plasmonic gain in current biased tilted {D}irac nodes},\
  }\href {https://doi.org/10.1038/s41467-022-35139-y} {\bibfield  {journal}
  {\bibinfo  {journal} {Nature Communications}\ }\textbf {\bibinfo {volume}
  {13}},\ \bibinfo {pages} {7667} (\bibinfo {year} {2022})}\BibitemShut
  {NoStop}%
\bibitem [{\citenamefont {Uykur}\ \emph {et~al.}(2019)\citenamefont {Uykur},
  \citenamefont {Li}, \citenamefont {Kuntscher},\ and\ \citenamefont
  {Dressel}}]{Uykur2019}%
  \BibitemOpen
  \bibfield  {author} {\bibinfo {author} {\bibfnamefont {E.}~\bibnamefont
  {Uykur}}, \bibinfo {author} {\bibfnamefont {W.}~\bibnamefont {Li}}, \bibinfo
  {author} {\bibfnamefont {C.~A.}\ \bibnamefont {Kuntscher}},\ and\ \bibinfo
  {author} {\bibfnamefont {M.}~\bibnamefont {Dressel}},\ }\bibfield  {title}
  {\bibinfo {title} {Optical signatures of energy gap in correlated dirac
  fermions},\ }\href {https://doi.org/10.1038/s41535-019-0158-z} {\bibfield
  {journal} {\bibinfo  {journal} {npj Quantum Materials}\ }\textbf {\bibinfo
  {volume} {4}},\ \bibinfo {pages} {19} (\bibinfo {year} {2019})}\BibitemShut
  {NoStop}%
\bibitem [{\citenamefont {Hirata}\ \emph {et~al.}(2016)\citenamefont {Hirata},
  \citenamefont {Ishikawa}, \citenamefont {Miyagawa}, \citenamefont {Tamura},
  \citenamefont {Berthier}, \citenamefont {Basko}, \citenamefont {Kobayashi},
  \citenamefont {Matsuno},\ and\ \citenamefont {Kanoda}}]{Hirata2016}%
  \BibitemOpen
  \bibfield  {author} {\bibinfo {author} {\bibfnamefont {M.}~\bibnamefont
  {Hirata}}, \bibinfo {author} {\bibfnamefont {K.}~\bibnamefont {Ishikawa}},
  \bibinfo {author} {\bibfnamefont {K.}~\bibnamefont {Miyagawa}}, \bibinfo
  {author} {\bibfnamefont {M.}~\bibnamefont {Tamura}}, \bibinfo {author}
  {\bibfnamefont {C.}~\bibnamefont {Berthier}}, \bibinfo {author}
  {\bibfnamefont {D.}~\bibnamefont {Basko}}, \bibinfo {author} {\bibfnamefont
  {A.}~\bibnamefont {Kobayashi}}, \bibinfo {author} {\bibfnamefont
  {G.}~\bibnamefont {Matsuno}},\ and\ \bibinfo {author} {\bibfnamefont
  {K.}~\bibnamefont {Kanoda}},\ }\bibfield  {title} {\bibinfo {title}
  {Observation of an anisotropic {D}irac cone reshaping and ferrimagnetic spin
  polarization in an organic conductor},\ }\href
  {https://doi.org/10.1038/ncomms12666} {\bibfield  {journal} {\bibinfo
  {journal} {Nature Communications}\ }\textbf {\bibinfo {volume} {7}},\
  \bibinfo {pages} {12666} (\bibinfo {year} {2016})}\BibitemShut {NoStop}%
\bibitem [{\citenamefont {Hirata}\ \emph {et~al.}(2017)\citenamefont {Hirata},
  \citenamefont {Ishikawa}, \citenamefont {Matsuno}, \citenamefont {Kobayashi},
  \citenamefont {Miyagawa}, \citenamefont {Tamura}, \citenamefont {Berthier},\
  and\ \citenamefont {Kanoda}}]{HirataAnomalous}%
  \BibitemOpen
  \bibfield  {author} {\bibinfo {author} {\bibfnamefont {M.}~\bibnamefont
  {Hirata}}, \bibinfo {author} {\bibfnamefont {K.}~\bibnamefont {Ishikawa}},
  \bibinfo {author} {\bibfnamefont {G.}~\bibnamefont {Matsuno}}, \bibinfo
  {author} {\bibfnamefont {A.}~\bibnamefont {Kobayashi}}, \bibinfo {author}
  {\bibfnamefont {K.}~\bibnamefont {Miyagawa}}, \bibinfo {author}
  {\bibfnamefont {M.}~\bibnamefont {Tamura}}, \bibinfo {author} {\bibfnamefont
  {C.}~\bibnamefont {Berthier}},\ and\ \bibinfo {author} {\bibfnamefont
  {K.}~\bibnamefont {Kanoda}},\ }\bibfield  {title} {\bibinfo {title}
  {Anomalous spin correlations and excitonic instability of interacting 2{D}
  {W}eyl fermions},\ }\href {https://doi.org/10.1126/science.aan5351}
  {\bibfield  {journal} {\bibinfo  {journal} {Science}\ }\textbf {\bibinfo
  {volume} {358}},\ \bibinfo {pages} {1403} (\bibinfo {year}
  {2017})}\BibitemShut {NoStop}%
\bibitem [{\citenamefont {Ohki}\ \emph {et~al.}(2020)\citenamefont {Ohki},
  \citenamefont {Hirata}, \citenamefont {Tani}, \citenamefont {Kanoda},\ and\
  \citenamefont {Kobayashi}}]{OhkiChiral}%
  \BibitemOpen
  \bibfield  {author} {\bibinfo {author} {\bibfnamefont {D.}~\bibnamefont
  {Ohki}}, \bibinfo {author} {\bibfnamefont {M.}~\bibnamefont {Hirata}},
  \bibinfo {author} {\bibfnamefont {T.}~\bibnamefont {Tani}}, \bibinfo {author}
  {\bibfnamefont {K.}~\bibnamefont {Kanoda}},\ and\ \bibinfo {author}
  {\bibfnamefont {A.}~\bibnamefont {Kobayashi}},\ }\bibfield  {title} {\bibinfo
  {title} {Chiral excitonic instability of two-dimensional tilted {D}irac
  cones},\ }\href {https://doi.org/10.1103/PhysRevResearch.2.033479} {\bibfield
   {journal} {\bibinfo  {journal} {Phys. Rev. Res.}\ }\textbf {\bibinfo
  {volume} {2}},\ \bibinfo {pages} {033479} (\bibinfo {year}
  {2020})}\BibitemShut {NoStop}%
\bibitem [{\citenamefont {Ohki}\ \emph {et~al.}(2023)\citenamefont {Ohki},
  \citenamefont {Yoshimi}, \citenamefont {Kobayashi},\ and\ \citenamefont
  {Misawa}}]{OhkiGap}%
  \BibitemOpen
  \bibfield  {author} {\bibinfo {author} {\bibfnamefont {D.}~\bibnamefont
  {Ohki}}, \bibinfo {author} {\bibfnamefont {K.}~\bibnamefont {Yoshimi}},
  \bibinfo {author} {\bibfnamefont {A.}~\bibnamefont {Kobayashi}},\ and\
  \bibinfo {author} {\bibfnamefont {T.}~\bibnamefont {Misawa}},\ }\bibfield
  {title} {\bibinfo {title} {Gap opening mechanism for correlated dirac
  electrons in organic compounds
  {$\ensuremath{\alpha}\text{\ensuremath{-}}{(\mathrm{BEDT}\text{\ensuremath{-}}\mathrm{TTF})}_{2}{\mathrm{I}}_{3}$}
  and
  {$\ensuremath{\alpha}\text{\ensuremath{-}}{(\mathrm{BEDT}\text{\ensuremath{-}}\mathrm{TSeF})}_{2}{\mathrm{I}}_{3}$}},\
  }\href {https://doi.org/10.1103/PhysRevB.107.L041108} {\bibfield  {journal}
  {\bibinfo  {journal} {Phys. Rev. B}\ }\textbf {\bibinfo {volume} {107}},\
  \bibinfo {pages} {L041108} (\bibinfo {year} {2023})}\BibitemShut {NoStop}%
\bibitem [{\citenamefont {Kobayashi}\ \emph {et~al.}(2007)\citenamefont
  {Kobayashi}, \citenamefont {Katayama}, \citenamefont {Suzumura},\ and\
  \citenamefont {Fukuyama}}]{MesslessFermions}%
  \BibitemOpen
  \bibfield  {author} {\bibinfo {author} {\bibfnamefont {A.}~\bibnamefont
  {Kobayashi}}, \bibinfo {author} {\bibfnamefont {S.}~\bibnamefont {Katayama}},
  \bibinfo {author} {\bibfnamefont {Y.}~\bibnamefont {Suzumura}},\ and\
  \bibinfo {author} {\bibfnamefont {H.}~\bibnamefont {Fukuyama}},\ }\bibfield
  {title} {\bibinfo {title} {Massless fermions in organic conductor},\ }\href
  {https://doi.org/10.1143/JPSJ.76.034711} {\bibfield  {journal} {\bibinfo
  {journal} {Journal of the Physical Society of Japan}\ }\textbf {\bibinfo
  {volume} {76}},\ \bibinfo {pages} {034711} (\bibinfo {year}
  {2007})}\BibitemShut {NoStop}%
\bibitem [{\citenamefont {Osada}\ and\ \citenamefont
  {Kiswandhi}(2020)}]{OsadaCurrentInduced}%
  \BibitemOpen
  \bibfield  {author} {\bibinfo {author} {\bibfnamefont {T.}~\bibnamefont
  {Osada}}\ and\ \bibinfo {author} {\bibfnamefont {A.}~\bibnamefont
  {Kiswandhi}},\ }\bibfield  {title} {\bibinfo {title} {Possible
  current-induced phenomena and domain control in an organic {D}irac fermion
  system with weak charge ordering},\ }\href
  {https://doi.org/10.7566/JPSJ.89.103701} {\bibfield  {journal} {\bibinfo
  {journal} {Journal of the Physical Society of Japan}\ }\textbf {\bibinfo
  {volume} {89}},\ \bibinfo {pages} {103701} (\bibinfo {year}
  {2020})}\BibitemShut {NoStop}%
\bibitem [{\citenamefont {Yoshimura}\ \emph {et~al.}(2021)\citenamefont
  {Yoshimura}, \citenamefont {Sato},\ and\ \citenamefont {Osada}}]{osadaExp}%
  \BibitemOpen
  \bibfield  {author} {\bibinfo {author} {\bibfnamefont {K.}~\bibnamefont
  {Yoshimura}}, \bibinfo {author} {\bibfnamefont {M.}~\bibnamefont {Sato}},\
  and\ \bibinfo {author} {\bibfnamefont {T.}~\bibnamefont {Osada}},\ }\bibfield
   {title} {\bibinfo {title} {Experimental confirmation of massive {D}irac
  fermions in weak charge-ordering state in {$\alpha$-(BEDT-TTF)$_2$I$_3$}},\
  }\href {https://doi.org/10.7566/JPSJ.90.033701} {\bibfield  {journal}
  {\bibinfo  {journal} {Journal of the Physical Society of Japan}\ }\textbf
  {\bibinfo {volume} {90}},\ \bibinfo {pages} {033701} (\bibinfo {year}
  {2021})}\BibitemShut {NoStop}%
\bibitem [{\citenamefont {Wang}\ \emph {et~al.}(2019)\citenamefont {Wang},
  \citenamefont {L{\"u}}, \citenamefont {Wang}, \citenamefont {Feng},\ and\
  \citenamefont {Zheng}}]{Wang2019}%
  \BibitemOpen
  \bibfield  {author} {\bibinfo {author} {\bibfnamefont {Z.-Q.}\ \bibnamefont
  {Wang}}, \bibinfo {author} {\bibfnamefont {T.-Y.}\ \bibnamefont {L{\"u}}},
  \bibinfo {author} {\bibfnamefont {H.-Q.}\ \bibnamefont {Wang}}, \bibinfo
  {author} {\bibfnamefont {Y.~P.}\ \bibnamefont {Feng}},\ and\ \bibinfo
  {author} {\bibfnamefont {J.-C.}\ \bibnamefont {Zheng}},\ }\bibfield  {title}
  {\bibinfo {title} {Band gap opening in 8-{$Pmmn$} borophene by
  hydrogenation},\ }\href {https://doi.org/10.1021/acsaelm.9b00017} {\bibfield
  {journal} {\bibinfo  {journal} {ACS Applied Electronic Materials}\ }\textbf
  {\bibinfo {volume} {1}},\ \bibinfo {pages} {667} (\bibinfo {year}
  {2019})}\BibitemShut {NoStop}%
\bibitem [{\citenamefont {Yekta}\ \emph {et~al.}(2023)\citenamefont {Yekta},
  \citenamefont {Hadipour},\ and\ \citenamefont {Jafari}}]{Yekta2023}%
  \BibitemOpen
  \bibfield  {author} {\bibinfo {author} {\bibfnamefont {Y.}~\bibnamefont
  {Yekta}}, \bibinfo {author} {\bibfnamefont {H.}~\bibnamefont {Hadipour}},\
  and\ \bibinfo {author} {\bibfnamefont {S.~A.}\ \bibnamefont {Jafari}},\
  }\bibfield  {title} {\bibinfo {title} {Tunning the tilt of the {D}irac cone
  by atomic manipulations in 8{P}mmn borophene},\ }\href
  {https://doi.org/10.1038/s42005-023-01161-9} {\bibfield  {journal} {\bibinfo
  {journal} {Communications Physics}\ }\textbf {\bibinfo {volume} {6}},\
  \bibinfo {pages} {46} (\bibinfo {year} {2023})}\BibitemShut {NoStop}%
\bibitem [{\citenamefont {Holcomb}\ \emph {et~al.}(1993)\citenamefont
  {Holcomb}, \citenamefont {Collman},\ and\ \citenamefont {Little}}]{Holcomb}%
  \BibitemOpen
  \bibfield  {author} {\bibinfo {author} {\bibfnamefont {M.~J.}\ \bibnamefont
  {Holcomb}}, \bibinfo {author} {\bibfnamefont {J.~P.}\ \bibnamefont
  {Collman}},\ and\ \bibinfo {author} {\bibfnamefont {W.~A.}\ \bibnamefont
  {Little}},\ }\bibfield  {title} {\bibinfo {title} {Thermal difference
  spectroscopy},\ }\href {https://doi.org/10.1063/1.1143969} {\bibfield
  {journal} {\bibinfo  {journal} {Review of Scientific Instruments}\ }\textbf
  {\bibinfo {volume} {64}},\ \bibinfo {pages} {1862} (\bibinfo {year}
  {1993})}\BibitemShut {NoStop}%
\bibitem [{\citenamefont {Holcomb}\ \emph {et~al.}(1994)\citenamefont
  {Holcomb}, \citenamefont {Collman},\ and\ \citenamefont
  {Little}}]{Holcomb1994}%
  \BibitemOpen
  \bibfield  {author} {\bibinfo {author} {\bibfnamefont {M.~J.}\ \bibnamefont
  {Holcomb}}, \bibinfo {author} {\bibfnamefont {J.~P.}\ \bibnamefont
  {Collman}},\ and\ \bibinfo {author} {\bibfnamefont {W.~A.}\ \bibnamefont
  {Little}},\ }\bibfield  {title} {\bibinfo {title} {Optical evidence of an
  electronic contribution to the pairing interaction in superconducting
  {${\mathrm{Tl}}_{2}$${\mathrm{Ba}}_{2}$${\mathrm{Ca}}_{2}$
  ${\mathrm{Cu}}_{3}$${\mathrm{O}}_{10}$}},\ }\href
  {https://doi.org/10.1103/PhysRevLett.73.2360} {\bibfield  {journal} {\bibinfo
   {journal} {Phys. Rev. Lett.}\ }\textbf {\bibinfo {volume} {73}},\ \bibinfo
  {pages} {2360} (\bibinfo {year} {1994})}\BibitemShut {NoStop}%
\bibitem [{\citenamefont {Holcomb}\ \emph {et~al.}(1996)\citenamefont
  {Holcomb}, \citenamefont {Perry}, \citenamefont {Collman},\ and\
  \citenamefont {Little}}]{Holcomb1996}%
  \BibitemOpen
  \bibfield  {author} {\bibinfo {author} {\bibfnamefont {M.~J.}\ \bibnamefont
  {Holcomb}}, \bibinfo {author} {\bibfnamefont {C.~L.}\ \bibnamefont {Perry}},
  \bibinfo {author} {\bibfnamefont {J.~P.}\ \bibnamefont {Collman}},\ and\
  \bibinfo {author} {\bibfnamefont {W.~A.}\ \bibnamefont {Little}},\ }\bibfield
   {title} {\bibinfo {title} {Thermal-difference reflectance spectroscopy of
  the high-temperature cuprate superconductors},\ }\href
  {https://doi.org/10.1103/PhysRevB.53.6734} {\bibfield  {journal} {\bibinfo
  {journal} {Phys. Rev. B}\ }\textbf {\bibinfo {volume} {53}},\ \bibinfo
  {pages} {6734} (\bibinfo {year} {1996})}\BibitemShut {NoStop}%
\bibitem [{\citenamefont {Maldague}(1978)}]{maldague}%
  \BibitemOpen
  \bibfield  {author} {\bibinfo {author} {\bibfnamefont {P.~F.}\ \bibnamefont
  {Maldague}},\ }\bibfield  {title} {\bibinfo {title} {Many-body corrections to
  the polarizability of the two-dimensional electron gas},\ }\href
  {https://doi.org/https://doi.org/10.1016/0039-6028(78)90507-1} {\bibfield
  {journal} {\bibinfo  {journal} {Surface Science}\ }\textbf {\bibinfo {volume}
  {73}},\ \bibinfo {pages} {296} (\bibinfo {year} {1978})}\BibitemShut
  {NoStop}%
\bibitem [{\citenamefont {Nandy}\ and\ \citenamefont
  {Pesin}(2022)}]{sciPostWTe2}%
  \BibitemOpen
  \bibfield  {author} {\bibinfo {author} {\bibfnamefont {S.}~\bibnamefont
  {Nandy}}\ and\ \bibinfo {author} {\bibfnamefont {D.~A.}\ \bibnamefont
  {Pesin}},\ }\bibfield  {title} {\bibinfo {title} {Low-energy effective theory
  and anomalous hall effect in monolayer {$\mathrm{WTe}_2$}},\ }\href
  {https://doi.org/10.21468/SciPostPhys.12.4.120} {\bibfield  {journal}
  {\bibinfo  {journal} {SciPost Phys.}\ }\textbf {\bibinfo {volume} {12}},\
  \bibinfo {pages} {120} (\bibinfo {year} {2022})}\BibitemShut {NoStop}%
\bibitem [{\citenamefont {Rostami}\ and\ \citenamefont {Juri\ifmmode
  \check{c}\else \v{c}\fi{}i\ifmmode~\acute{c}\else
  \'{c}\fi{}}(2020)}]{juriNLHE}%
  \BibitemOpen
  \bibfield  {author} {\bibinfo {author} {\bibfnamefont {H.}~\bibnamefont
  {Rostami}}\ and\ \bibinfo {author} {\bibfnamefont {V.}~\bibnamefont
  {Juri\ifmmode \check{c}\else \v{c}\fi{}i\ifmmode~\acute{c}\else
  \'{c}\fi{}}},\ }\bibfield  {title} {\bibinfo {title} {Probing quantum
  criticality using nonlinear {H}all effect in a metallic {D}irac system},\
  }\href {https://doi.org/10.1103/PhysRevResearch.2.013069} {\bibfield
  {journal} {\bibinfo  {journal} {Phys. Rev. Res.}\ }\textbf {\bibinfo {volume}
  {2}},\ \bibinfo {pages} {013069} (\bibinfo {year} {2020})}\BibitemShut
  {NoStop}%
\bibitem [{\citenamefont {Du}\ \emph {et~al.}(2019)\citenamefont {Du},
  \citenamefont {Wang}, \citenamefont {Li}, \citenamefont {Lu},\ and\
  \citenamefont {Xie}}]{disorderNLHE}%
  \BibitemOpen
  \bibfield  {author} {\bibinfo {author} {\bibfnamefont {Z.~Z.}\ \bibnamefont
  {Du}}, \bibinfo {author} {\bibfnamefont {C.~M.}\ \bibnamefont {Wang}},
  \bibinfo {author} {\bibfnamefont {S.}~\bibnamefont {Li}}, \bibinfo {author}
  {\bibfnamefont {H.-Z.}\ \bibnamefont {Lu}},\ and\ \bibinfo {author}
  {\bibfnamefont {X.~C.}\ \bibnamefont {Xie}},\ }\bibfield  {title} {\bibinfo
  {title} {Disorder-induced nonlinear {H}all effect with time-reversal
  symmetry},\ }\href {https://doi.org/10.1038/s41467-019-10941-3} {\bibfield
  {journal} {\bibinfo  {journal} {Nature Communications}\ }\textbf {\bibinfo
  {volume} {10}},\ \bibinfo {pages} {3047} (\bibinfo {year}
  {2019})}\BibitemShut {NoStop}%
\bibitem [{\citenamefont {Du}\ \emph {et~al.}(2018)\citenamefont {Du},
  \citenamefont {Wang}, \citenamefont {Lu},\ and\ \citenamefont
  {Xie}}]{Du2018}%
  \BibitemOpen
  \bibfield  {author} {\bibinfo {author} {\bibfnamefont {Z.~Z.}\ \bibnamefont
  {Du}}, \bibinfo {author} {\bibfnamefont {C.~M.}\ \bibnamefont {Wang}},
  \bibinfo {author} {\bibfnamefont {H.-Z.}\ \bibnamefont {Lu}},\ and\ \bibinfo
  {author} {\bibfnamefont {X.~C.}\ \bibnamefont {Xie}},\ }\bibfield  {title}
  {\bibinfo {title} {Band signatures for strong nonlinear {H}all effect in
  bilayer {${\mathrm{WTe}}_{2}$}},\ }\href
  {https://doi.org/10.1103/PhysRevLett.121.266601} {\bibfield  {journal}
  {\bibinfo  {journal} {Phys. Rev. Lett.}\ }\textbf {\bibinfo {volume} {121}},\
  \bibinfo {pages} {266601} (\bibinfo {year} {2018})}\BibitemShut {NoStop}%
\bibitem [{\citenamefont {Lahiri}\ \emph {et~al.}(2022)\citenamefont {Lahiri},
  \citenamefont {Das}, \citenamefont {Culcer},\ and\ \citenamefont
  {Agarwal}}]{QMDCulcer1}%
  \BibitemOpen
  \bibfield  {author} {\bibinfo {author} {\bibfnamefont {S.}~\bibnamefont
  {Lahiri}}, \bibinfo {author} {\bibfnamefont {K.}~\bibnamefont {Das}},
  \bibinfo {author} {\bibfnamefont {D.}~\bibnamefont {Culcer}},\ and\ \bibinfo
  {author} {\bibfnamefont {A.}~\bibnamefont {Agarwal}},\ }\bibfield  {title}
  {\bibinfo {title} {Intrinsic nonlinear conductivity induced by the quantum
  metric dipole},\ }\href {https://arxiv.org/abs/2207.02178} {\bibfield
  {journal} {\bibinfo  {journal} {arXiv preprint arXiv:2207.02178}\ } (\bibinfo
  {year} {2022})}\BibitemShut {NoStop}%
\bibitem [{\citenamefont {Bhalla}\ \emph {et~al.}(2022)\citenamefont {Bhalla},
  \citenamefont {Das}, \citenamefont {Culcer},\ and\ \citenamefont
  {Agarwal}}]{QMDCulcer2}%
  \BibitemOpen
  \bibfield  {author} {\bibinfo {author} {\bibfnamefont {P.}~\bibnamefont
  {Bhalla}}, \bibinfo {author} {\bibfnamefont {K.}~\bibnamefont {Das}},
  \bibinfo {author} {\bibfnamefont {D.}~\bibnamefont {Culcer}},\ and\ \bibinfo
  {author} {\bibfnamefont {A.}~\bibnamefont {Agarwal}},\ }\bibfield  {title}
  {\bibinfo {title} {Resonant second-harmonic generation as a probe of quantum
  geometry},\ }\href {https://doi.org/10.1103/PhysRevLett.129.227401}
  {\bibfield  {journal} {\bibinfo  {journal} {Phys. Rev. Lett.}\ }\textbf
  {\bibinfo {volume} {129}},\ \bibinfo {pages} {227401} (\bibinfo {year}
  {2022})}\BibitemShut {NoStop}%
\bibitem [{\citenamefont {Zhou}\ \emph {et~al.}(2022)\citenamefont {Zhou},
  \citenamefont {Zhang}, \citenamefont {Yu}, \citenamefont {Zhu},\ and\
  \citenamefont {Su}}]{NLThermal}%
  \BibitemOpen
  \bibfield  {author} {\bibinfo {author} {\bibfnamefont {D.-K.}\ \bibnamefont
  {Zhou}}, \bibinfo {author} {\bibfnamefont {Z.-F.}\ \bibnamefont {Zhang}},
  \bibinfo {author} {\bibfnamefont {X.-Q.}\ \bibnamefont {Yu}}, \bibinfo
  {author} {\bibfnamefont {Z.-G.}\ \bibnamefont {Zhu}},\ and\ \bibinfo {author}
  {\bibfnamefont {G.}~\bibnamefont {Su}},\ }\bibfield  {title} {\bibinfo
  {title} {Fundamental distinction between intrinsic and extrinsic nonlinear
  thermal {H}all effects},\ }\href
  {https://doi.org/10.1103/PhysRevB.105.L201103} {\bibfield  {journal}
  {\bibinfo  {journal} {Phys. Rev. B}\ }\textbf {\bibinfo {volume} {105}},\
  \bibinfo {pages} {L201103} (\bibinfo {year} {2022})}\BibitemShut {NoStop}%
\bibitem [{\citenamefont {Gorbar}\ \emph {et~al.}(2002)\citenamefont {Gorbar},
  \citenamefont {Gusynin}, \citenamefont {Miransky},\ and\ \citenamefont
  {Shovkovy}}]{gorbarLi2}%
  \BibitemOpen
  \bibfield  {author} {\bibinfo {author} {\bibfnamefont {E.~V.}\ \bibnamefont
  {Gorbar}}, \bibinfo {author} {\bibfnamefont {V.~P.}\ \bibnamefont {Gusynin}},
  \bibinfo {author} {\bibfnamefont {V.~A.}\ \bibnamefont {Miransky}},\ and\
  \bibinfo {author} {\bibfnamefont {I.~A.}\ \bibnamefont {Shovkovy}},\
  }\bibfield  {title} {\bibinfo {title} {Magnetic field driven metal-insulator
  phase transition in planar systems},\ }\href
  {https://doi.org/10.1103/PhysRevB.66.045108} {\bibfield  {journal} {\bibinfo
  {journal} {Phys. Rev. B}\ }\textbf {\bibinfo {volume} {66}},\ \bibinfo
  {pages} {045108} (\bibinfo {year} {2002})}\BibitemShut {NoStop}%
\bibitem [{\citenamefont {Iurov}\ \emph {et~al.}(2017)\citenamefont {Iurov},
  \citenamefont {Gumbs}, \citenamefont {Huang},\ and\ \citenamefont
  {Balakrishnan}}]{gumbs2017}%
  \BibitemOpen
  \bibfield  {author} {\bibinfo {author} {\bibfnamefont {A.}~\bibnamefont
  {Iurov}}, \bibinfo {author} {\bibfnamefont {G.}~\bibnamefont {Gumbs}},
  \bibinfo {author} {\bibfnamefont {D.}~\bibnamefont {Huang}},\ and\ \bibinfo
  {author} {\bibfnamefont {G.}~\bibnamefont {Balakrishnan}},\ }\bibfield
  {title} {\bibinfo {title} {Thermal plasmons controlled by different
  thermal-convolution paths in tunable extrinsic {D}irac structures},\ }\href
  {https://doi.org/10.1103/PhysRevB.96.245403} {\bibfield  {journal} {\bibinfo
  {journal} {Phys. Rev. B}\ }\textbf {\bibinfo {volume} {96}},\ \bibinfo
  {pages} {245403} (\bibinfo {year} {2017})}\BibitemShut {NoStop}%
\bibitem [{\citenamefont {Tsaran}\ \emph {et~al.}(2017)\citenamefont {Tsaran},
  \citenamefont {Kavokin}, \citenamefont {Sharapov}, \citenamefont {Varlamov},\
  and\ \citenamefont {Gusynin}}]{spikes}%
  \BibitemOpen
  \bibfield  {author} {\bibinfo {author} {\bibfnamefont {V.~Y.}\ \bibnamefont
  {Tsaran}}, \bibinfo {author} {\bibfnamefont {A.~V.}\ \bibnamefont {Kavokin}},
  \bibinfo {author} {\bibfnamefont {S.~G.}\ \bibnamefont {Sharapov}}, \bibinfo
  {author} {\bibfnamefont {A.~A.}\ \bibnamefont {Varlamov}},\ and\ \bibinfo
  {author} {\bibfnamefont {V.~P.}\ \bibnamefont {Gusynin}},\ }\bibfield
  {title} {\bibinfo {title} {Entropy spikes as a signature of {L}ifshitz
  transitions in the {D}irac materials},\ }\href
  {https://doi.org/10.1038/s41598-017-10643-0} {\bibfield  {journal} {\bibinfo
  {journal} {Scientific Reports}\ }\textbf {\bibinfo {volume} {7}},\ \bibinfo
  {pages} {10271} (\bibinfo {year} {2017})}\BibitemShut {NoStop}%
\bibitem [{\citenamefont {Gradshteyn}\ \emph {et~al.}(1988)\citenamefont
  {Gradshteyn}, \citenamefont {Ryzhik},\ and\ \citenamefont
  {Romer}}]{gradshteynLi2}%
  \BibitemOpen
  \bibfield  {author} {\bibinfo {author} {\bibfnamefont {I.~S.}\ \bibnamefont
  {Gradshteyn}}, \bibinfo {author} {\bibfnamefont {I.~M.}\ \bibnamefont
  {Ryzhik}},\ and\ \bibinfo {author} {\bibfnamefont {R.~H.}\ \bibnamefont
  {Romer}},\ }\bibfield  {title} {\bibinfo {title} {Tables of integrals,
  series, and products},\ }\href {https://doi.org/10.1119/1.15756} {\bibfield
  {journal} {\bibinfo  {journal} {American Journal of Physics}\ }\textbf
  {\bibinfo {volume} {56}},\ \bibinfo {pages} {958} (\bibinfo {year}
  {1988})}\BibitemShut {NoStop}%
\end{thebibliography}%
\end{document}